\documentclass[aps,pre,twocolumn,showpacs,superscriptaddress]{revtex4-2}

\usepackage{graphicx}
\usepackage{dcolumn}
\usepackage{bm}
\usepackage{amsmath,amssymb,mathtools}
\usepackage{color}
\usepackage{appendix}
\usepackage{xcolor}

\usepackage[para,online,flushleft]{threeparttable}
\usepackage{booktabs}
\usepackage{multirow}

\definecolor{darkblue}{rgb}{0,0,0.6}
\definecolor{darkred}{rgb}{0.6,0,0}
\definecolor{darkgreen}{rgb}{0,0.6,0}

\usepackage[colorlinks=true,urlcolor=darkblue,citecolor=darkblue,linkcolor=darkred,hyperfootnotes=false]{hyperref}

\usepackage{setspace}
\usepackage{appendix}

\makeatother

\def\bibsection{\section*{\refname}}

\begin{document}

\title{Phase separation of chemokinetic active particles}

\author{Euijoon Kwon}
\thanks{These authors contributed equally.}
\affiliation{Department of Physics and Astronomy \& Center for Theoretical Physics, Seoul National University, Seoul 08826, Republic of Korea}

\author{Yongjae Oh}
\thanks{These authors contributed equally.}
\affiliation{Department of Physics and Astronomy \& Center for Theoretical Physics, Seoul National University, Seoul 08826, Republic of Korea}

\author{Yongjoo Baek}
\email{y.baek@snu.ac.kr}
\affiliation{Department of Physics and Astronomy \& Center for Theoretical Physics, Seoul National University, Seoul 08826, Republic of Korea}

\date{\today}
 
\begin{abstract}
Motility-induced phase separation (MIPS) is a well-studied nonequilibrium collective phenomenon observed in active particles. Recently, there has been growing interest in how coupling the self-propulsion of active particles to chemical degrees of freedom affects MIPS. Although the effects of chemotaxis on MIPS have been extensively studied, little is known about how chemokinesis affects MIPS. In this study, we demonstrate that various patterns can be induced when active particles consume chemicals and exhibit chemokinesis, where higher chemical concentrations enhance self-propulsion without causing alignment with the chemical gradient. We discover that MIPS is intensified if chemical consumption is proportional to particle density (as in the basal metabolic regime), but it is suppressed if chemical consumption is closely tied to particle motion (as in the active metabolic regime). While the former produces large-scale phase separation via coarsening, the latter suppresses the coarsening process, leading to microphase separation and oscillating patterns. We also derive a hydrodynamic theory that describes these findings.
\end{abstract}


\maketitle

\section*{Introduction}
\label{sec:intro}

Active matter, or substances consisting of self-propelled units, has attracted much attention as a novel class of nonequilibrium systems exhibiting various collective phenomena~\cite{RamaswamyARCMP2010, MarchettiRMP2013, BechingerRMP2016, RamaswamyJSM2017, JulicherRPP2018, GompperJPCM2020, BowickPRX2022, Vrugt2024}. Among these is the motility-induced phase separation (MIPS), which refers to the spontaneous separation between high- and low-density regions stemming from self-propulsion and short-range repulsion of active particles~\cite{Tailleur2008, Fily2012, Redner2013, Bialke2013, Cates2015}. Various theoretical frameworks have been developed to describe the phenomena, including top-down derivations of active field theories~\cite{Wittkowski2014, Tiribocchi2015, Tjhung2018} and bottom-up derivations of coarse-grained hydrodynamic models~\cite{Bertin2006, Bialke2013}. Predictions of these theories have been tested by extensive simulations~\cite{Stenhammar2013, Stenhammar2014, Caporusso2020, Shi2020}.




For simplicity, these studies have focused on active particles that propel themselves at constant force and change directions only on their own accord. However, there have also been studies of how the coupling between active particles and chemicals affects the collective phenomena. They have mainly focused on chemotaxis, \textit{i.e.}, the tendency of microorganisms or particles to move up (chemoattraction) or down (chemorepulsion) the chemical concentration gradient. At the hydrodynamic level, such effects have been modeled as the parallel or antiparallel alignment of local flux~\cite{Keller1970,Keller1971,Meyer2014,Zhao2023} or polarization~\cite{Liebchen2015, Liebchen2017} with the concentration gradient. If the chemicals are released by the particles themselves (autochemotaxis), it is well known that chemoattraction leads to complete phase separation~\cite{Keller1970, Keller1971, Meyer2014}, while chemorepulsion generates a variety of patterns, such as finite-size dynamic clusters and traveling waves~\cite{Liebchen2015, Liebchen2017}. Moreover, when externally injected chemoattractants are taken up by active particles, MIPS can be arrested or create oscillating patterns~\cite{Zhao2023}. On the other hand, when chemotaxis arises as a consequence of the diffusiophoresis of spherical particles, the coupling between active particles and chemicals has been shown to induce both complete phase separation and dynamic clustering, where multiple small clusters coexist with particles constantly joining and leaving them~\cite{Pohl2014dynamic,Pohl2015selfphoretic,Fadda2023interplay}. Dynamic clustering of the platinum--gold Janus particles, empirically observed in \cite{Theurkauff2012}, was attributed to this effect. These results demonstrate that the interplay between active particles and chemicals can induce a host of pattern-formation behaviors.


However, there is another type of coupling between active particles and chemicals. For example, while the direction of self-propulsion is unaffected by the chemical concentration gradient, the magnitude of self-propulsion may change depending on the concentration level. This phenomenon, called chemokinesis, occurs in various kinds of cells including sperm cells~\cite{Ralt1994, Inamdar2007}, neural cells~\cite{Richards2004}, leukocytes~\cite{Wilkinson1984, Wilkinson1990}, and bacteria~\cite{Brown1993, Armitage1997, Barbara2003, Garren2014, Chen2015} as they respond to external chemical stimuli. While chemokinesis often coexists with chemotaxis in microorganisms, different types of receptors are responsible for chemokinetic and chemotactic responses, implying that the strengths of those responses can be controlled separately~\cite{Leaman2021}. Indeed, there are both naturally occurring and artificially induced examples of cells which exhibit strong chemokinesis but very little chemotaxis, such as neutrophils in aged human subjects~\cite{Sapey2014} and human endothelial cells after the CMG2 gene is blocked~\cite{Cryan2022}. There are also synthetic active particles known to exhibit chemokinetic response to their fuel with no or little chemotaxis. These include platinum--polystyrene Janus particles propelled by diffusiophoresis~\cite{Howse2007} and platinum--gold rods propelled by electrophoresis~\cite{Moran2021}, which are both fueled by hydrogen peroxide in water. For the Janus particles driven by diffusiophoresis, analyses show that chemotaxis is negligible when the mobility constants of the two halves of the particle are not very different~\cite{Pohl2014dynamic,Pohl2015selfphoretic,Popescu2018}.


Then, how does chemokinesis affect the pattern formation of active matter? The question has been addressed mainly in the context of how active particles respond to an externally imposed chemical concentration gradient. For example, a model study predicted that chemokinesis mixed with chemotaxis enhances the migratory response of bacteria to an imposed chemoattractant concentration gradient compared to those only capable of chemotaxis~\cite{Jakuszeit2021}. In experiments with self-propelled rods with an externally imposed fuel concentration gradient, chemokinesis was found to be responsible for the accumulation of particles in the low-concentration region~\cite{Moran2021}. Broadening the scope to include light-activated particles, analogous effects can also be found in photokinetic bacteria, which have been used to control their self-organized patterns and transport~\cite{Frangipane2018, Arlt2018, Arlt2019, Koumakis2019}.


But what if the active particles create the concentration gradients as they take up the chemicals? Little has been studied about how chemokinesis in the absence of chemotaxis affects the phase-separation and pattern-formation behaviors of autonomous systems formed by active particles and chemicals. To address this scenario, we explore a model system in which active particles are subject to chemokinesis and chemical consumption. We observe that chemokinetic effects are sufficient to induce a rich range of pattern formation behaviors, which can be described analytically through a hydrodynamic model and linear stability analysis.




\begin{figure*}
	\includegraphics[width=0.99\textwidth]{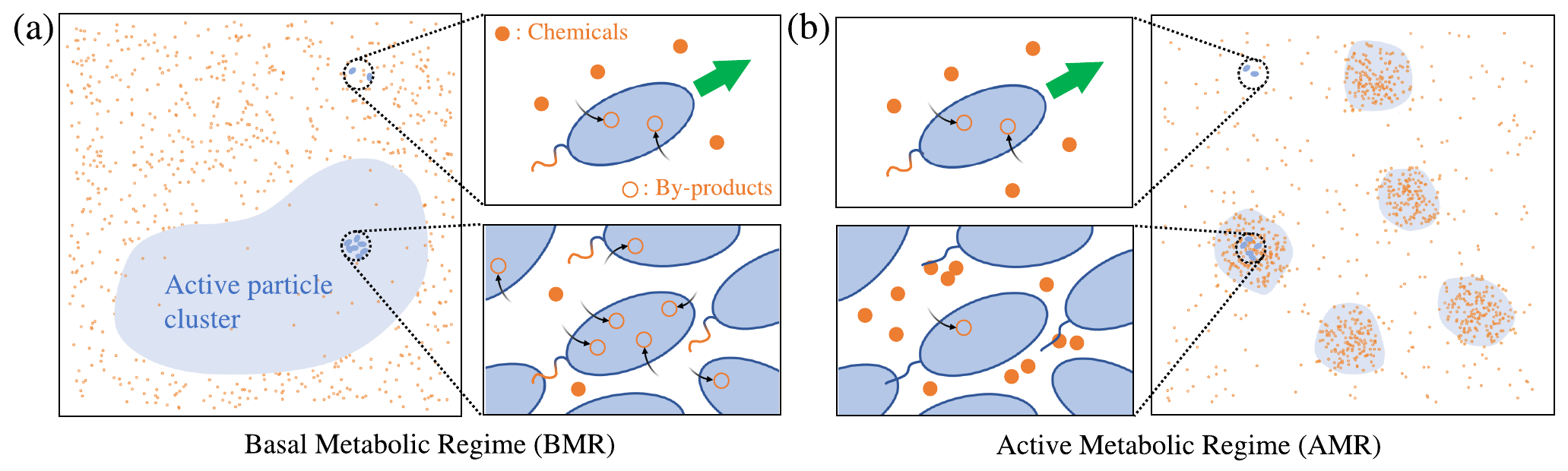}
    \caption{\label{fig:BMR_AMR_schematic} \textbf{Schematic illustrations of two different regimes of local chemical consumption.} Each blue elliptical particle with a tail represents an active particle, whose direction of self-propulsion is indicated by a green arrow. Regions colored in light blue correspond to active particle clusters. Chemicals fueling the active particles are represented by filled orange dots, while by-products produced by chemical consumption are depicted with empty orange dots.
    (a) In the Basal Metabolic Regime, where chemical consumption rate is proportional to particle density, chemical concentration tends to be higher where particle density is lower. This enhances phase separation, creating a single large-scale cluster.
    (b) In the Active Metabolic Regime, where chemical consumption rate is determined by particle flux, chemical concentration tends to be lower where particle density is higher. This suppresses phase separation, creating multiple finite-size clusters.}
\end{figure*}

\section*{Methods}
\label{sec:model}

\subsection*{Particle-based model}


To formulate on a concrete basis how chemokinesis and chemical consumption affect the dynamics of active particles, we first introduce a particle-based model, from which a hydrodynamic theory can be derived by coarse-graining. More specifically, we consider $N$ overdamped active Brownian particles (ABPs) moving within a two-dimensional chemical concentration profile $n(\mathbf{r},t)$. The equation of motion for each ABP is given by
\begin{align}
    \label{eq:eom}
    \dot{\mathbf{r}}_k (t) &= \tilde{v}(n(\mathbf{r}_k,t))\,\mathbf{e}_k - \sum_{k'\neq k}\mu \bm\nabla_k V(|\mathbf{r}_k-\mathbf{r}_{k'}|)\nonumber\\
    &\quad+ \sqrt{2\mu T}\, \bm\xi_k
    \;,
\end{align}
where $\mathbf{r}_k$ denotes the position of the $k$th ABP, $\tilde{v}$ the magnitude of self-propulsion, $\mu$ the mobility of each ABP, $\mathbf{e}_k$ the direction of self-propulsion exhibiting angular Brownian motion characterized by the rotational diffusion coefficient $D_\mathrm{r}$, $V$ the pairwise interaction between ABPs, $T$ the temperature of the medium, and $\bm\xi_k$ a Gaussian white noise satisfying $\langle {\bm\xi}_k (t) \rangle =0$ and $\langle {\bm\xi}_k (t) {\bm\xi}_{k'} ^\textsf{T} (t') \rangle =\delta_{kk'} \mathbb{I} \delta(t-t')$ for the $2\times 2$ identity matrix $\mathbb{I}$. Note that chemokinesis is incorporated into the model via the dependence of $\tilde{v}$ on the local chemical concentration $n(\mathbf{r}_k,t)$. Unlike chemotaxis, which aligns or anti-aligns $\mathbf{e}_k$ to the concentration gradient, chemokinesis modifies only the motility of the active particle, as captured by the function $\tilde{v}(n)$. For simplicity, we focus on the linear regime where $\tilde{v}(n)=\alpha n$, with $\alpha>0$ controlling the strength of chemokinesis.

The chemokinetic effect directly captured by this model is called orthokinesis. We note that there is also klinokinesis, which couples $D_\mathrm{r}$ to the chemical concentration. While the distinction between these two is crucial in the biological context, they are not fundamentally different in the physical context since a klinokinetic system can be mapped to an orthokinetic system by suitable rescaling of time. Thus, here we focus only on the latter.


We also model how the chemical is consumed by active particles. To prevent the complete depletion of the chemical and keep the system out of equilibrium, we assume that the chemical is uniformly supplied into the system with a constant injection rate $I$. Then, the time evolution of the concentration profile can be written as
\begin{align}
    \label{eq:eom_chem}
    \dot{n}(\mathbf{r},t) = D_\mathrm{c} \nabla^2 n + I - \lambda f(n,\{\mathbf{r}_k,\dot{\mathbf{r}}_k\})
    \;,
\end{align}
where $D_\mathrm{c}$ is the diffusion constant of the chemical, $\lambda$ is the consumption rate, and $f$ is a function governing how the chemical consumption depends on the state of all ABPs and the chemical. If $\lambda = 0$, the particle dynamics is decoupled from the chemical, which we call the \emph{vanilla} model. The behaviors of $f$ have been empirically studied for various species under different physical conditions~\cite{Videler1990, Ohlberger2007}. For simplicity, here we focus on two regimes of $f$ that would differently affect the phase separation behavior of ABPs.

(i) {\em Basal Metabolic Regime} (BMR). In this case, the chemical is mostly used to maintain the routine metabolism of the ABPs, not their motion. Then the chemical consumption rate solely depends on how many ABPs are around, as expressed by
\begin{align} \label{eq:BMR_f}
    f(n,\{\mathbf{r}_k,\dot{\mathbf{r}}_k\}) = n(\mathbf{r}) \sum_{k'=1} ^N \delta (\mathbf{r} - \mathbf{r}_{k'})
    \;.
\end{align}

In the BMR, the chemical consumption rate increases with the density of ABPs, which would lower the chemical concentration within ABP clusters. The resulting chemokinetic slowdown hinders the ABPs from escaping the cluster. Thus, one can naturally expect that the phase-separating tendencies are strengthened in the BMR, as schematically illustrated in Fig.~\ref{fig:BMR_AMR_schematic}(a).

(ii) {\em Active Metabolic Regime} (AMR).
In this regime, the chemical consumption is tightly coupled to the motion of the ABPs, as expressed by
\begin{align} \label{eq:AMR_f}
    f(n,\{\mathbf{r}_k,\dot{\mathbf{r}}_k\}) = n(\mathbf{r}) \sum_{k'=1} ^N  \dot{\mathbf{r}}_{k'} \cdot \mathbf{e}_{k'} \delta (\mathbf{r} - \mathbf{r}_{k'})
    \;.
\end{align}
Note that the chemical is consumed when the self-propulsion orientation $\mathbf{e}_k$ and the velocity $\dot{\mathbf{r}}_{k}$ are in the same direction, while any motion of the particle against the self-propulsion leads to the chemical production. This is a natural consequence of the mechanochemical coupling obeying the Onsager reciprocal relations, as discussed in \cite{RamaswamyJSM2017,gaspard2018fluctuating,dadhichi2018origins,dabelow2019irreversibility,crosato2019irreversibility,oh2023effects}.


In the AMR, the chemical consumption rate depends on both the particle density and the effective speed along the self-propulsion direction. If the particle density is sufficiently high within the ABP clusters formed by MIPS, the decrease of particle speed overwhelms the increase of particle density, reducing the chemical consumption and raising the chemical concentration within the ABP clusters. Then the ensuing chemokinetic speedup makes the cluster more prone to disintegration. Thus, in contrast to the BMR, one can expect that the phase-separating tendencies may be weakened in the AMR, as schematically illustrated in Fig.~\ref{fig:BMR_AMR_schematic}(b).

To investigate the effects of chemical consumption and chemokinesis on phase separation and pattern formation, we directly simulate the particle-based model following Eqs.~\eqref{eq:eom}--\eqref{eq:AMR_f}, with the short-range repulsion between particles given by the Weeks--Chandler--Andersen (WCA) potential
\begin{align}
    V(r) =
    \begin{cases}
        4 u_0 \left[\left(\frac{\sigma}{r}\right)^{12}-\left(\frac{\sigma}{r}\right)^{6}\right]+u_0 & \text{for $r < 2^{1/6}\sigma$} \\
        0 &\text{for $r\geq 2^{1/6}\sigma$}
    \end{cases}\;, \label{eq:wca_potential}
\end{align}
where $u_0$ sets the magnitude of the potential, and $\sigma$ is the effective diameter of each particle. We fix $u_0 = 100$ and $\sigma = 1$, which, together with the persistence time fixed at $\tau_\mathrm{p} \equiv D_\mathrm{r} ^{-1} = 1/3$, define the natural units of length, time, and energy, respectively. We note that the Dirac delta function appearing in Eqs.~\eqref{eq:BMR_f} and \eqref{eq:AMR_f} are discretized in the simulation, so that its value becomes $1$ only when the particle is in the grid corresponding to $\mathbf{r}$ and $0$ otherwise. The simulation results are discussed in the Results section.

\begin{figure*}
\includegraphics[width=0.99\textwidth]{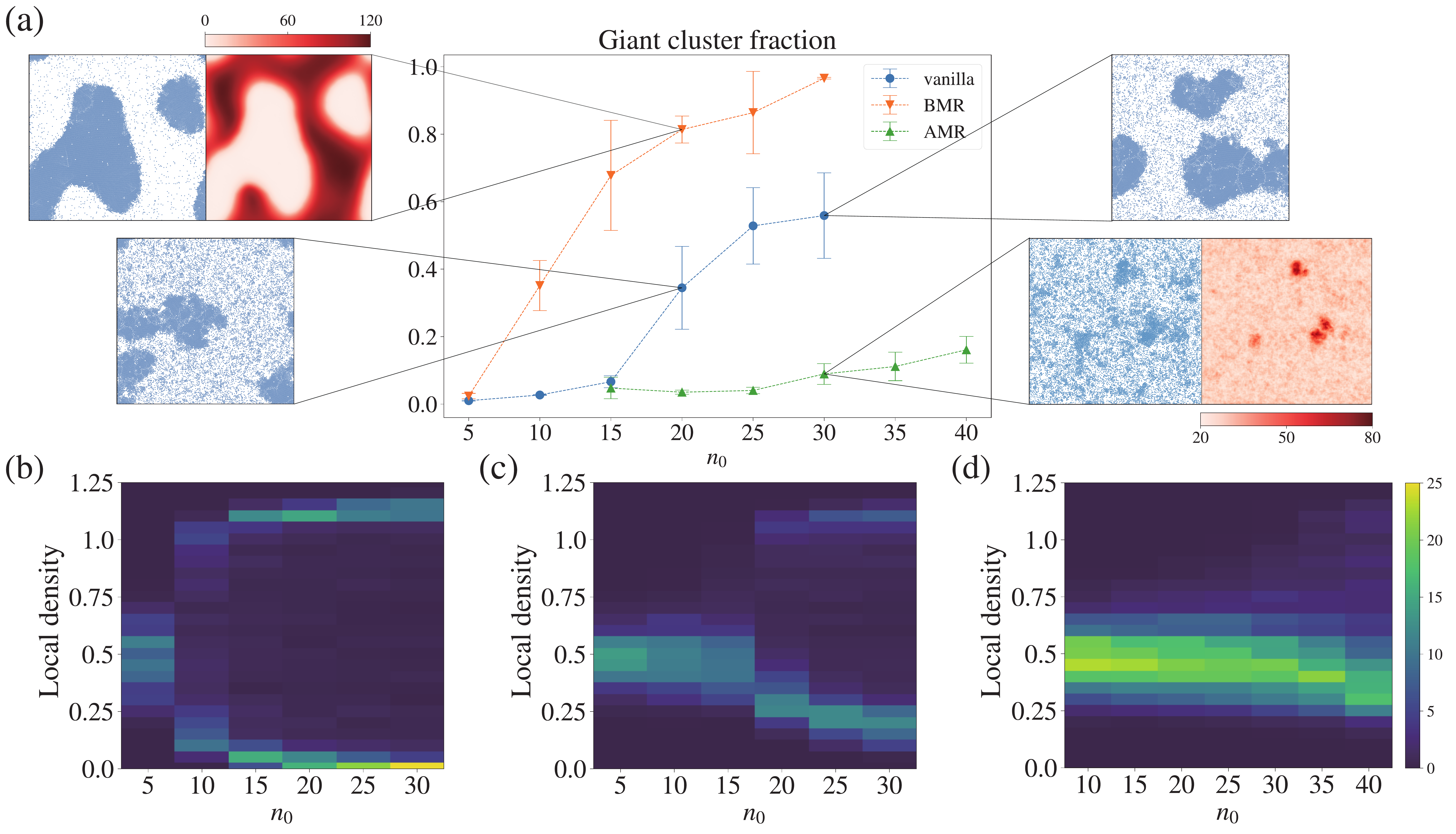}
    \caption{
\textbf{Phase separation behaviors of the particle-based model with repulsive interactions.}
(a) Fraction of particles belonging to the largest cluster (identified using the DBSCAN algorithm~\cite{Ester1996density}) as the global chemical concentration $n_0$ is varied.
Error bars indicate standard deviations obtained using six different realizations. Insets: Snapshots displaying the locations of particles (blue dots) and the heatmaps of chemical concentration (red shades), which are sampled from the four selected data points of the main figure. Note that chemical concentration is irrelevant and thus not shown for the vanilla model. The system is a two-dimensional torus of side length $L = 200$ and global particle density $\rho_0 = 1/2$. We also use $\alpha = 1$, $D_0 \equiv \mu T= 1$, $D_\mathrm{r} = 3$, $\sigma = 1$, $u_0 = 100$, $\lambda = 0.1$, and $D_\mathrm{c} = 1$. See the discussions below Eqs.~\eqref{eq:eom}, \eqref{eq:eom_chem}, and \eqref{eq:wca_potential} for descriptions of the parameters. We also show the histograms of local particle density as $n_0$ is varied for (b) the Basal Metabolic Regime (BMR), (c) the vanilla model, and (d) the Active Metabolic Regie (AMR). All physical quantities and snapshots are measured at time $t = 200$. See Supplementary Movie 1A (1B) for an animated comparison between the vanilla model and the BMR (AMR) at $n_0 = 20$ ($n_0 = 30$).}
\label{fig:WCA_gcf_local_density}
\end{figure*}

\subsection*{Hydrodynamic model}
\label{sec:LSA}

To understand the phase separation and pattern formation behaviors exhibited by the particle-based model, we also propose a system of hydrodynamic equations for the particle density $\rho(\mathbf{r},t)$ and the chemical concentration $n(\mathbf{r},t)$, which can be derived from the particle-based model assuming slow spatial variations of the fields and fast angular relaxation of the ABPs. See Appendix~\ref{sec:appendixa} for the detailed derivation, which closely follows the procedure given in Ref.~\cite{Bialke2013}. The resulting hydrodynamic equations are given by
\begin{align} \label{eq:hydro}
 \nonumber
    \dot{\rho}(\mathbf{r},t)&= \nabla \cdot \left\{ D\nabla \rho + \frac{v(\rho,n)}{2D_\mathrm{r}} \nabla [v(\rho,n) \rho]-\kappa \nabla(\nabla^2 \rho) \right\}\;,
    \\
     \dot{n}(\mathbf{r},t)&=D_\mathrm{c} \nabla ^2 n + I - \begin{cases}
\lambda \rho n & \text{(BMR)}\\
\lambda \rho v(\rho, n)n &\text{(AMR)}
\end{cases}
\;.
\end{align}
Here $D$ is the ABP diffusion coefficient, $\kappa$ is a parameter setting the interfacial energy, and $v(\rho,n)$ is the effective self-propulsion strength accounting for both acceleration by chemokinesis and deceleration by interparticle repulsion. Through the rest of the study, we focus on the linear regime where
\begin{align} \label{eq:v_rhon}
v(\rho,n) = \alpha n -\rho \zeta\:,
\end{align}
so that the positive coefficient $\zeta$ quantifies the slowdown effect. We propose the hydrodynamic model expressed by Eqs.~\eqref{eq:hydro} and \eqref{eq:v_rhon} as a minimal model for exploring the effects of chemokinesis and chemical consumption on MIPS.

\section*{Results}

\subsection*{Behaviors of the particle-based model}

To verify whether the BMR (AMR) facilitates (suppresses) MIPS, we let the particle-based model described in the Methods section evolve from a homogeneous initial state to $t = 200$ and observe its phase separation behaviors. In Fig.~\ref{fig:WCA_gcf_local_density}(a), we show the fraction of particles belonging to the largest cluster (also called the giant cluster) for the BMR, the AMR, and the vanilla model using the same set of parameters (except for $\lambda = 0$ used in the vanilla model). The BMR develops a larger particle cluster compared to the vanilla model, with the chemical concentration becoming higher outside the particle cluster. In contrast, the AMR strongly suppresses the growth of clusters, with the chemical concentration becoming higher inside the particle cluster. We highlight that the giant clusters formed by the BMR have smooth interfaces, while those formed by the AMR tends to be of rugged shapes. This rules out the presence of the Ostwald process driven by the positive interfacial tension in the AMR, which is crucial for the large-scale phase separation in the BMR.

The same conclusion can also be drawn from the histograms of the local particle density for the BMR [see Fig.~\ref{fig:WCA_gcf_local_density}(b)], the vanilla model [see Fig.~\ref{fig:WCA_gcf_local_density}(c)], and the AMR [see Fig.~\ref{fig:WCA_gcf_local_density}(d)] at different values of $n_0$. For the BMR, the separation between dense and sparse regions occurs at the smallest value of $n_0$ among the three cases, and the density gap at a fixed value of $n_0$ is also the largest. For the AMR, on the contrary, the phase separation can hardly be detected by the histograms. In other words, the AMR does not exhibit phase separation, or the dense particle clusters are too small to be captured by the histograms.


These results indicate that the BMR facilitates the large-scale MIPS, whereas the AMR suppresses MIPS and allows only microphase separation. As discussed earlier, the latter effect is attributable to increased self-propulsion of the ABPs inside the high-density particle clusters. In more realistic situations, the self-propulsion of ABPs may eventually saturate to an upper bound instead of growing linearly with the chemical concentration as in Eq.~\eqref{eq:v_rhon}. In this case, we observe that the suppression of MIPS in the AMR becomes weaker, with the particle clusters becoming larger as the upper bound on self-propulsion becomes tighter. This lends further support to our view that increased self-propulsion suppresses MIPS. See Appendix~\ref{sec:AppendixC} and Fig.~\ref{fig:tanh} for more detailed discussions.

\subsection*{Behaviors of the hydrodynamic model}

\subsubsection*{Linear stability analysis} \label{sec:app_LSA}

The hydrodynamic model expressed by Eqs.~\eqref{eq:hydro} and \eqref{eq:v_rhon} allows us to predict the phase-transition behaviors of the system through linear stability analysis (LSA) of the homogeneous solution. We begin by noting that the homogeneous density and concentration profiles $\rho(\mathbf{r},t) = \rho_0$ and $n(\mathbf{r},t)=n_0$ can solve Eq.~\eqref{eq:hydro} when the injection rate is given by
\begin{align}
\label{eq:injection}
I = \begin{cases}
    \lambda \rho_0 n_0 &\text{(BMR)}\\
    \lambda \rho_0 v(\rho_0,n_0)n_0 &\text{(AMR)}
\end{cases}\;.
\end{align}
Assuming this condition, we let the density and the concentration profiles fluctuate by $\delta\rho(\mathbf{r},t)$ and $\delta n(\mathbf{r},t)$, respectively. Taking the Ansatz $\delta \rho = \delta \rho_0 e^{i\mathbf{q}\cdot \mathbf{r} + \omega t}$ and $\delta n = \delta n_0 e^{i\mathbf{q}\cdot \mathbf{r} + \omega t}$, Eq.~\eqref{eq:hydro} can be linearized as an eigenvalue equation
\begin{align} \label{eq:eig_eq}
    \begin{pmatrix} \mathcal{A} & \mathcal{B} \\ \mathcal{C} & \mathcal{D} \end{pmatrix} \begin{pmatrix} \delta \rho_0 \\ \delta n_0 \end{pmatrix} =  \omega \begin{pmatrix} \delta \rho_0 \\ \delta n_0 \end{pmatrix}
    \;.
\end{align}
See Appendix~\ref{sec:AppendixB} for explicit expressions of the matrix elements $\mathcal{A}$, $\mathcal{B}$, $\mathcal{C}$, and $\mathcal{D}$. The eigenvalue $\omega$ thus satisfies the characteristic equation $\omega ^2 - (\mathcal{A}+\mathcal{D})\omega + \mathcal{A}\mathcal{D} - \mathcal{B}\mathcal{C}=0$, whose roots are given by
\begin{align} \label{eq:omega}
    \omega_{\pm} = \frac{1}{2} \left[ \mathcal{A} + \mathcal{D} \pm \sqrt{(\mathcal{A}-\mathcal{D})^2 + 4\mathcal{B}\mathcal{C}} \right]
    \;.
\end{align}

Depending on how the real and the imaginary parts of $\omega_\pm$ vary with $q^2 \equiv \mathbf{q} \cdot \mathbf{q}$, we can identify three phases. First, when $\mathrm{Re}\,\omega_{\pm} < 0$ for all $q^2$, the homogeneous profiles are stable against fluctuations, resulting in the homogeneous (H) phase. Second, if $\mathrm{Re}\,\omega_{\pm} > 0$ for some interval of $q^2$ and $\mathrm{Im}\,\omega_{\pm} = 0$ there, then any slight deviations from the homogeneous profiles would grow exponentially until the LSA breaks down. Assuming that the lack of oscillation persists beyond the LSA regime, this behavior would result in a coexistence phase where high-density and low-density regions form a largely stable interface, which we call the stationary (S) phase. Finally, when there is an interval of $q^2$ with $\mathrm{Re}\,\omega_{\pm} > 0$ and $\mathrm{Im}\,\omega_{\pm} \neq 0$, perturbations of the homogeneous profiles exhibit oscillations with growing amplitudes. We call this regime the oscillatory (O) phase. See Fig.~\ref{fig:appendix_AMR_LSA} and Appendix~\ref{sec:AppendixB} for detailed derivations of the phase boundaries.

\begin{figure*}
	\includegraphics[width=0.99\textwidth]{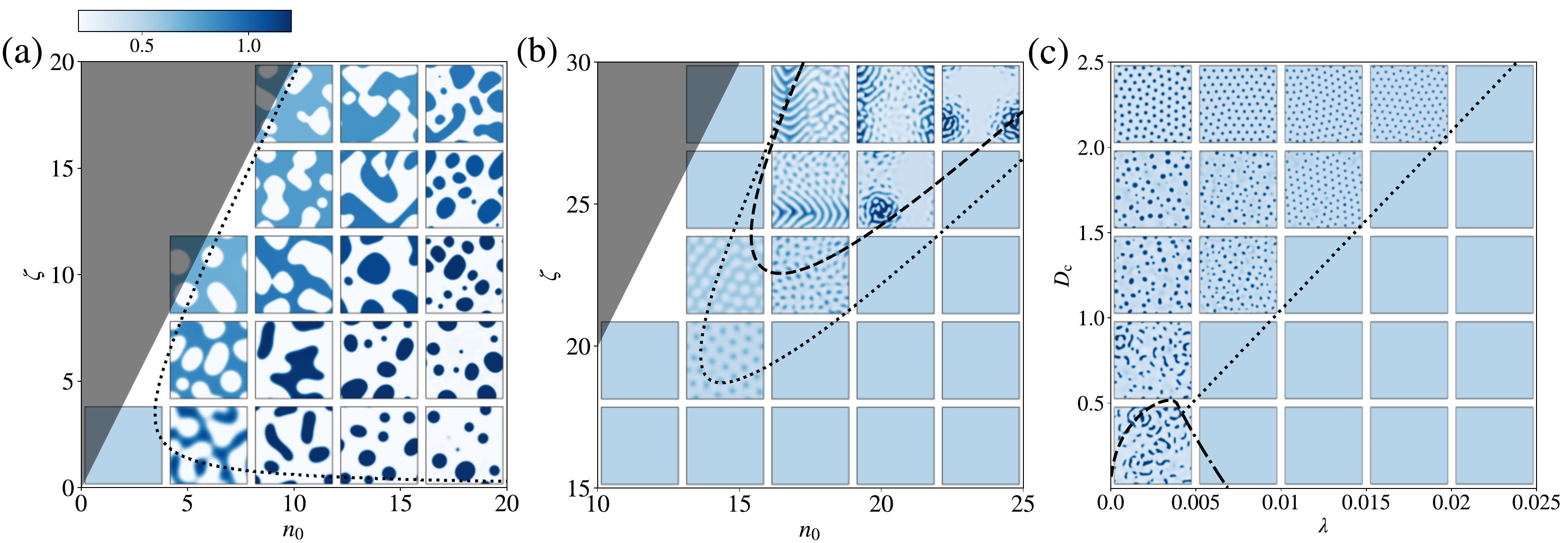} \caption{\label{fig:hydro_phase_diagrams} \textbf{Snapshots and phase diagrams of the hydrodynamic model.} These are shown for (a) the Basal Metabolic Regime with $\lambda = 0.5$, $D_\mathrm{c} = 0.1$ and for the Active Metabolic Regime with (b) $\lambda = 0.01$, $D_\mathrm{c} = 1$ and (c) $n_0 = 20$, $\zeta = 22$. Note that $n_0$ denotes the global chemical concentration, $\lambda$ the chemical consumption rate, $D_\mathrm{c}$ the diffusion constant of the chemical, and $\zeta$ the coefficient describing the deceleration due to interparticle interactions. Predictions of linear stability analysis for the phase boundaries are indicated by the dotted (H--S boundary), the dash-dotted (H--O boundary), and the dashed (S--O boundary) lines. For comparison, snapshots of particle distribution (with darker regions indicating higher density) obtained by hydrodynamic simulations are shown together. The values of $\lambda$ and $D_\mathrm{c}$ for (a) and (b), and $\lambda$ and $D_\mathrm{c}$ for (c) in each snapshot correspond to the coordinates of its center. The unphysical region with $v_\mathrm{eff} \equiv \alpha n_0 - \rho_0 \zeta < 0$ is grayed out for (a) and (b). For this and all the following results in this paper, we use $\alpha = 1$, $\rho_0=1/2$, $D=1$, $D_\mathrm{r} = 3$, and $\kappa = 20$, unless mentioned otherwise. See Supplementary Movies 2A, 2B, and 2C for animated versions of (a), (b), and (c), respectively.}
\end{figure*}

Now, based on the LSA described so far, we investigate various phases of the hydrodynamic model. In the following, we separately discuss the LSA predictions for the phase boundaries in the BMR and in the AMR, which are then compared with simulation results.

\subsubsection*{BMR facilitates phase separation}

For the BMR, one can easily check that $\mathcal{B}\mathcal{C} >0$ for all $q^2$ as long as $v_\mathrm{eff} \equiv \alpha n_0 - \rho_0 \zeta >0$, which must be true in the regime where Eq.~\eqref{eq:v_rhon} is a good approximation. Thus, the BMR lacks the O phase, and we only need to consider the boundary between the H and the S phases. As detailed in Appendix~B, the boundary is derived as
\begin{align} \label{eq:BMR_phase}
    D_{\text{BMR}} = \frac{\rho_0 \zeta (\alpha n_0 - \rho_0 \zeta)}{D_\mathrm{r}}
    \;,
\end{align}
so that $D>D_{\text{BMR}}$ ($D<D_{\text{BMR}}$) corresponds to the H phase (S phase). Interestingly, $D_{\text{BMR}}$ does not depend on the value of the chemical consumption rate $\lambda$, marking the onset of MIPS whenever $\lambda > 0$.

When $\lambda = 0$, there is no chemical consumption, so $n(\mathbf{r},t)$ stays at its initial value $n_0$. Since the chemical concentration does not change anywhere, the effective self-propulsion strength can be written as $v(\rho,n) = v_0 - \rho\zeta$, where $v_0 \equiv \alpha n_0$ is a fixed constant. In this case, chemokinesis becomes completely irrelevant, effectively reducing the dynamics of the system to those of this vanilla model decoupled from the chemicals. Then, as detailed in Appendix~\ref{sec:AppendixB}, the phase boundary is determined by the effective diffusion coefficient
\begin{align}
    D_{\text{eff}} \equiv D + \frac{(v_0 - \rho_0 \zeta )(v_0 - 2\rho_0 \zeta)}{2D_\mathrm{r}}
\end{align}
that governs the collective relaxation of density profiles when the self-propulsion and the short-range repulsion between particles are taken into account (via $v_0$ and $\zeta$, respectively). MIPS occurs when and only when sufficiently strong deceleration effects (represented by $\zeta$) induce $D_{\text{eff}} < 0$, so that `negative diffusion' amplifies the density fluctuations. This yields the phase boundary
\begin{align} \label{eq:vanilla_phase}
    D_{\text{vanilla}} = \frac{(v_0 - \rho_0 \zeta )(2\rho_0 \zeta - v_0)}{2D_\mathrm{r}}
    \;,
\end{align}
with $D>D_{\text{vanilla}}$ ($D<D_{\text{vanilla}}$) corresponding to the H phase (S phase). Comparing the above formula with Eq.~\eqref{eq:BMR_phase}, it is clear that $D_\text{BMR} > D_\text{vanilla}$ always holds, indicating that nonzero $\lambda$ broadens the S phase at the expense of the H phase. Thus, as previously expected, the BMR facilitates MIPS.

We observe that the phase boundary changes discontinuously from $D_\text{vanilla}$ to $D_\text{BMR}$ as the chemical consumption is turned on. When the chemicals are irrelevant ($\lambda = 0$), MIPS requires that $D_\mathrm{eff}$ is negative, as discussed above. In contrast, when $\lambda > 0$, high local particle density and low local chemical concentration can support each other via positive feedback, which can amplify density and concentration modulations even as diffusive currents governed by positive $D_\mathrm{eff}$ try to suppress them. For this reason, the mechanism of MIPS for $\lambda > 0$ (which may happen even for $D_\mathrm{eff} > 0$) is fundamentally different from the corresponding mechanism for $\lambda = 0$ (which requires $D_\mathrm{eff} < 0$), which explains the discontinuous change of the phase boundary.

To check the validity of the LSA, we perform numerical simulations of Eq.~\eqref{eq:hydro} via the pseudo-spectral method.
In Fig.~\ref{fig:hydro_phase_diagrams}(a), we show the phase boundary (indicated by a dotted curve) predicted by the LSA together with snapshots of the numerical simulations at $t = 2\times10^3$ as the global mean concentration $n_0$ and the deceleration parameter $\zeta$ are varied within the physical regime (the grayed-out area corresponds to the unphysical regime where $v(\rho,n) < 0$). The two control parameters are chosen to represent the two major factors contributing to MIPS, namely the self-propulsion $\alpha n_0$ and the deceleration $\rho_0 \zeta$. Note that the coordinates of the center of each snapshot indicate the associated values of $n_0$ and $\zeta$. Clearly, the predicted boundary correctly separates the homogeneous snapshot from the phase-separated ones. 

We also note that, unlike the vanilla MIPS, chemokinetic effects allow the particles to form clusters (for high $n_0$) or bubbles (for low $n_0$) even at the same value of $\rho_0$. This phenomenon, along with demonstrations of coarsening dynamics, is presented in more detail in Fig.~\ref{fig:hydro_BMR_time_evol}, Supplementary Movies~5A, 5B, and Appendix~\ref{sec:AppendixD}.

\begin{figure*}
	\includegraphics[width=0.99\textwidth]{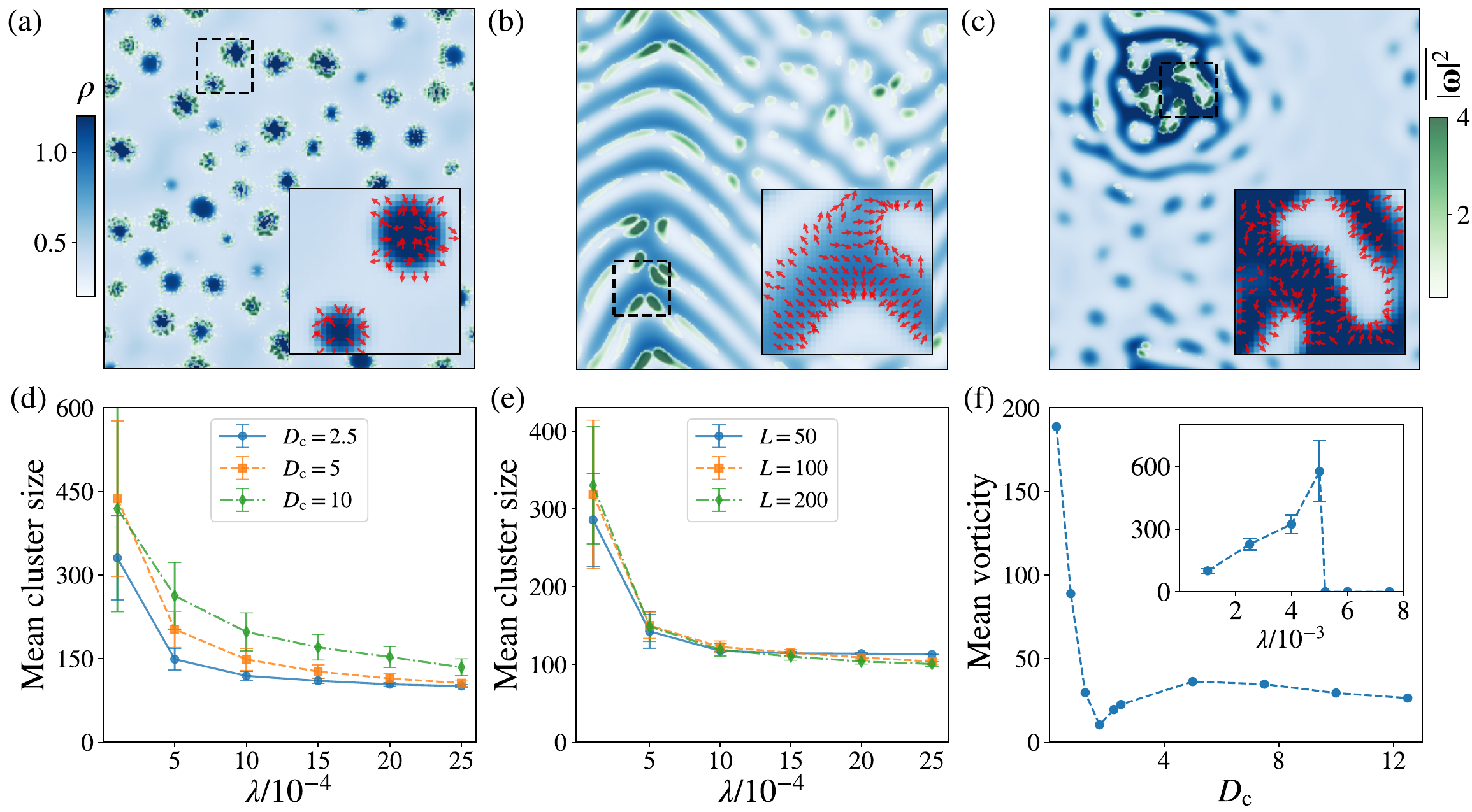}
    \caption{\label{fig:hydro_MSC_Vor} 
    \textbf{Phase separation behaviors of the hydrodynamic model in the Active Metabolic Regime.} Particle density $\rho(\mathbf{r},t)$ (blue) and the square magnitude of local vorticity $|\boldsymbol\omega(\mathbf{r},t)|^2$ (green), rescaled by its standard deviation, are shown for (a) diffusive droplets ($\lambda = 0.0025$, $D_\mathrm{c} = 1.75$, $n_0=10$, $\zeta = 22$), (b) ballistic bands ($\lambda = 0.01$, $D_\mathrm{c}=1$, $n_0=17.5$, $\zeta=28.5$), and (c) radiating ripples ($\lambda = 0.01$, $D_\mathrm{c}=1$, $n_0=23.5$, $\zeta=28.5$). Darker green color corresponds to the region with high local vorticity. Through (a--c), particle velocity fields are indicated by red arrows in the insets. (d, e) Mean cluster size in the S phase as the chemical consumption rate $\lambda$, the chemical diffusion coefficient $D_\mathrm{c}$, and the system size $L$ are varied. The error bars indicate the standard deviation of cluster size. (f) Mean vorticity as $D_\mathrm{c}$ is varied for $\lambda = 0.0025$ and (inset) the same quantity as $\lambda$ is varied for $D_\mathrm{c} = 0.25$. Note that $n_0 = 20$ and $\zeta = 22$ in (d--f). 
    }
\end{figure*}

\subsubsection*{AMR induces small, oscillating clusters}
\label{subsection:hydro_AMR}
The LSA for the AMR can be done similarly, but it differs from that for the BMR in that $\mathcal{B}$ and $\mathcal{C}$ in Eq.~\eqref{eq:eig_eq} may have different signs ($\mathcal{B} < 0$ and $\mathcal{C} > 0$). The eigenvalues $\omega_\pm$, obtained in Eq.~\eqref{eq:omega}, may then have nonzero imaginary parts. Thus, the AMR exhibits the O phase in addition to the H and the S phases, which leads to some extra complications in predicting the phase boundaries. Nonetheless, the LSA can be carried out exactly, as detailed in Appendix~B. The analysis yields a set of inequalities determining the range of $D_\mathrm{eff}$ belonging to the H phase, which is always broadened by the presence of the particle--chemical coupling ($\lambda > 0$) always broadens the H phase. This confirms our previous expectation that the AMR suppresses MIPS.

Examples of the $\zeta$--$n_0$ and the $D_\mathrm{c}$--$\lambda$ phase diagrams for the AMR are shown in Fig.~\ref{fig:hydro_phase_diagrams}(b) and (c), respectively. The former parameter set is chosen for direct comparison with Fig.~\ref{fig:hydro_phase_diagrams}(a), and the latter parameter set is chosen to represent how the strength of the particle--chemical coupling $\lambda$ and the time scale of the chemical dynamics $1/D_\mathrm{c}$ modify the phase separation behaviors. The LSA prediction for the H--S phase boundary is marked by the dotted line, the H--O boundary by the dash-dotted line, and the S--O boundary by the dashed line. For comparison, we also show the snapshots of the particle density obtained by numerical simulations of Eq.~\eqref{eq:hydro} via the pseudo-spectral method, where the coordinates of the center of each square corresponds to the parameters used in the simulation. The predicted H--S and the H--O boundaries, for which we expect the LSA to be exact, are in good agreement with the numerics. In contrast, the O phase seems to be broader than indicated by the predicted S--O boundary, as exemplified by the lower left corner of Fig.~\ref{fig:hydro_phase_diagrams}(c). Since the LSA is valid only up to linear order in density and concentration fluctuations, it is only natural that the true S--O boundary would deviate from the LSA prediction.

Besides the existence of new phase boundaries, the patterns shown by the AMR vastly differ from those of the BMR. In Fig.~\ref{fig:hydro_MSC_Vor}, we provide close-ups of the representative patterns generated by the AMR in the S and the O phases. Higher-density regions are indicated by darker blue shades, while green is used to show the mean vorticity $\boldsymbol{\omega} = \nabla \times \mathbf{u}$ of the particle flux, where $\mathbf{u} = \mathbf{J}/\rho$ is the local particle velocity defined in terms of the particle current density
\begin{align}
    \mathbf{J} = - \left\{ D\nabla \rho + \frac{v(\rho,n)}{2D_\mathrm{r}} \nabla [v(\rho,n) \rho]-\kappa \nabla(\nabla^2 \rho) \right\}
\end{align}
appearing in Eq.~\eqref{eq:hydro}. We note that high vorticity tends to coincide with regions where patterns evolve more rapidly. With these in mind, we can identify at last three distinct patterns, namely diffusive droplets [see Fig.~\ref{fig:hydro_MSC_Vor}(a)] well within the S phase, ballistic bands near the boundaries of the O phase [see Fig.~\ref{fig:hydro_MSC_Vor}(b)], and radiating ripples well within the O phase [see Fig.~\ref{fig:hydro_MSC_Vor}(c)]. It seems that the latter two patterns are essentially the same, the only difference being that radiating ripples have a smaller length scale. This is also corroborated by the LSA, which shows that the unstable oscillating fluctuations tend to exhibit shorter wavelengths in the regimes of radiating ripples compared to ballistic bands.

Since all these patterns exist beyond the regime where the LSA is valid, in the following we rely on the results of hydrodynamic simulations to examine how they depend on various control parameters.

Let us first look into the properties of the diffusive droplets. As indicated by Fig.~\ref{fig:hydro_MSC_Vor}(d), the average number of particles forming each droplet (which we call the \textit{mean cluster size}) decreases with $\lambda$ and increases with $D_\mathrm{c}$. The former is in agreement with our intuition that the particle--chemical coupling in the AMR suppresses MIPS. The latter is due to the high $D_\mathrm{c}$ strongly suppressing the short-wavelength undulations of the chemical concentration. Since all patterns are formed by the particle density and the chemical concentration sharing the same characteristic length scale, this effect also drives the particles to form larger droplets. However, as implied by the snapshots in Fig.~\ref{fig:hydro_phase_diagrams}(c), the mean cluster size starts to increase again when $D_\mathrm{c}$ decreases below $2.5$. This is due to the droplets becoming increasingly motile and actively colliding with each other as the system approaches the S--O boundary.


All these behaviors are in stark contrast to the case of the BMR, where MIPS eventually forms a single macroscopic particle cluster via coarsening. Indeed, Fig.~\ref{fig:hydro_MSC_Vor}(e) indicates that the mean cluster size does not change appreciably with the system size $L$ when $\lambda > 0$. In other words, the size of each droplet stays finite, and the S phase of the AMR exhibits microphase separation. This is also supported by the disintegration of a large particle cluster in the S phase, which is further discussed in Fig.~\ref{fig:AMR_time_evol}, Supplementary Movie 6, and Appendix~\ref{sec:AppendixD}.

Now, we examine how the system changes quantitatively as it transitions from the S phase with diffusive droplets to the O phase with ballistic bands and radiating ripples. As already shown in Figs.~\ref{fig:hydro_MSC_Vor}(a)--(c), the latter two patterns exhibit spatially extended regions of high vorticity, reflecting their large-scale evolution in time. In contrast, the high-vorticity regions for the diffusive droplets are limited to the droplet boundaries. This implies that the mean vorticity
\begin{align} \label{eq:mean_vorticity}
    \text{Mean vorticity} \equiv \frac{1}{L^2} \int d^2 \mathbf{r} \,\left|\nabla \times \mathbf{u}(\mathbf{r})\right|^2
\end{align}
can be used as an indicator of the O phase. Indeed, Fig.~\ref{fig:hydro_MSC_Vor}(f) shows that the mean vorticity dramatically increases as $D_\mathrm{c}$ is decreased for $\lambda = 0.0025$, crossing the S--O boundary shown in Fig.~\ref{fig:hydro_phase_diagrams}(c). Notably, even before reaching the S--O boundary, the mean vorticity gradually increases as $D_\mathrm{c}$ is decreased, which is then followed by a sharp dip around $D_\mathrm{c} = 2$. This nonmonotonic behavior might be due to the individual droplets becoming increasingly motile as $D_\mathrm{c}$ is reduced, which then merge together by collisions and shorten the boundaries, thus reducing the high-vorticity regions as well.

\begin{figure}
\includegraphics[width=\columnwidth]{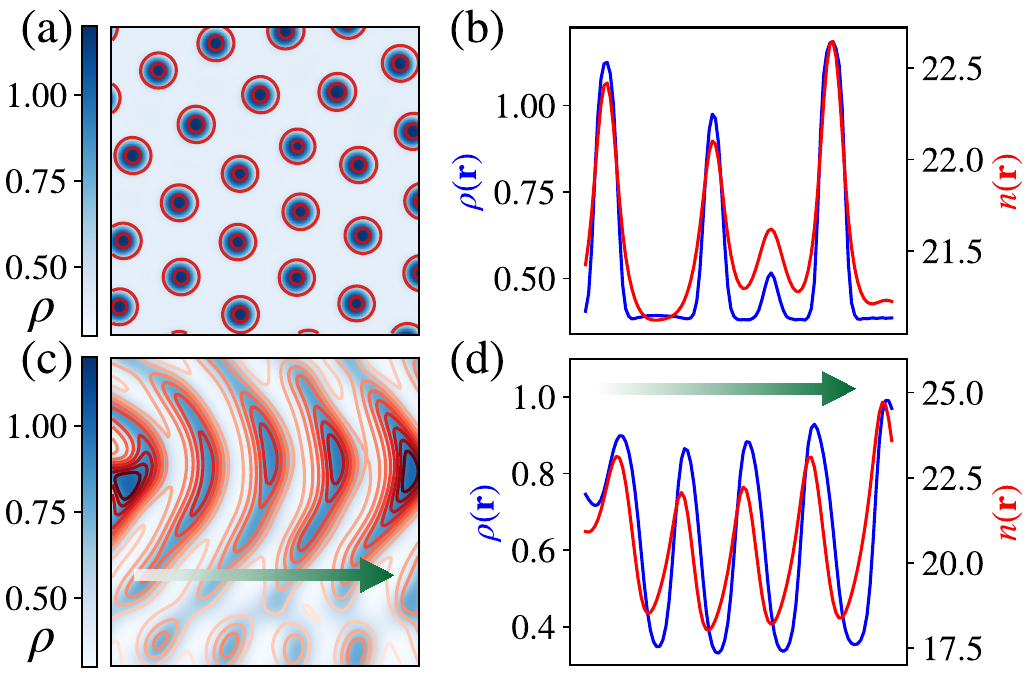}
    \caption{\label{fig:fig_hydro_osc_exp} 
    \textbf{Profiles of patterns formed by the hydrodynamic model in the Active Metabolic Regime}. Particle density $\rho(\mathbf{r})$ and chemical concentration $n(\mathbf{r})$ are shown for
    (a, b) the S phase with $D_\mathrm{c} = 2.25, \lambda = 0.0025, n_0 = 20, \zeta = 22$ and (c, d) the O phase with $D_\mathrm{c} = 1, \lambda = 0.01, n_0 = 25.5, \zeta = 17.5$. In (a) and (c), $\rho(\mathbf{r})$ is indicated by heatmaps (blue shades) $n(\mathbf{r})$ by contours, so that the values of $n(\mathbf{r})$ at adjacent contour lines differ by $0.8$. Note that (b) and (d) show cross sections of (a) and (c), respectively. The arrows in (c) and (d) indicate the directions in which the patterns move. See Supplementary Movie 3 for an animated version.}
\end{figure}

The inset of Fig.~\ref{fig:hydro_MSC_Vor}(f), which shows the mean vorticity for $D_\mathrm{c} = 0.25$ as $\lambda$ is varied, similarly captures the O phase lying between the S and the H phases in the lower part of Fig.~\ref{fig:hydro_phase_diagrams}(c). Notably, in apparent contrast to Fig.~\ref{fig:hydro_MSC_Vor}(f), the mean vorticity keeps increasing as $\lambda$ approaches the phase boundary. This is because the boundary crossed here is between the O and the H phases. The oscillating patterns, driven by the particle--chemical coupling, move more vigorously with increasing $\lambda$. However, when $\lambda$ increases beyond the H--O boundary, the patterns become transient and quickly die out. For this reason, the mean vorticity exhibits a sharp drop at the H--O boundary.

With the qualitative differences between different patterns thus established,
we look into how the particle--chemical coupling creates the traveling patterns in the O phase. When $D_\mathrm{c}$ is large, the chemical concentration is slaved to the particle density, so that their undulations overlap almost exactly as shown in Fig.~\ref{fig:fig_hydro_osc_exp}(a) and (b). This leads to the diffusive droplets observed in the S phase. In contrast, when $D_\mathrm{c}$ is sufficiently small, the particle and the chemical dynamics may have comparable time scales, in which case the following scenario becomes possible. Suppose that the particle density undulations are slightly shifted to the right with respect to those of the chemical concentration. Then, for each density peak, the chemical concentration is lower (higher) on the right (left), decreasing (increasing) the speed of particles there. This means that particles tend to accumulate on the right and evaporate on the left, making the peak travel to the right. The concentration peak follows the density peak but cannot catch up for a prolonged period of time since $D_\mathrm{c}$ is small. This is precisely what happens for the density and the concentration fields shown in Figs.~\ref{fig:fig_hydro_osc_exp}(c) and (d), which exhibit band-like patterns traveling to the right.

We stress that these traveling patterns in the O phase are consequences of the nonreciprocity reflected in the different signs of $\mathcal{B}$ and $\mathcal{C}$ appearing in Eq.~\eqref{eq:eig_eq}.
This is similar to how various multispecies active fluids with nonreciprocal interspecies interactions generate traveling patterns~\cite{you2020nonreciprocity, saha2020scalar, fruchart2021non, dinelli2023non-reciprocity, mandal2024robustness, pisegna2024emergent}.




The microphase separation in the AMR is consistent with the behaviors of the particle-based model. However, the distinction between the S and O phases of the AMR was not so pronounced in the mechanically interacting particle-based model. This may be because the microscopic noises, which are neglected in the hydrodynamic model, overwhelm the oscillating patterns in the particle-based model. This raises the question of whether the O phase of the AMR identified by the LSA is merely an artifact of the hydrodynamic model. Below, we address this issue using a different particle-based model, which is expected to have much lower noise than the model with WCA interactions.

\subsection*{Particle-based simulation of quorum-sensing ABPs} \label{sec:particle}


As a particle-based model more easily comparable with the hydrodynamic counterpart, here we introduce the \textit{quorum-sensing ABPs}. While their dynamics share Eqs.~\eqref{eq:eom_chem}--\eqref{eq:AMR_f} with the mechanically interacting ABPs, Eq.~\eqref{eq:eom} is replaced with
\begin{align}
    \dot{\mathbf{r}}_k (t) = v\big(\varrho_\mathrm{loc}(\mathbf{r}_k, t), n(\mathbf{r}_k,t)\big) \mathbf{e}_k + \sqrt{2\mu T}\, \bm{\xi}_k\;, \label{eq:eom_qs}    
\end{align}
where the magnitude of self-propulsion is given by 
\begin{align}
    v(\varrho_{\mathrm{loc}},n)=\max\left[\alpha n - \varrho_{\mathrm{loc}}\zeta,\,0\right] \label{eq:v_qs}
\end{align}
for the local particle density evaluated as
\begin{align} \label{eq:loc_den}
    \varrho_{\text{loc}} (\mathbf{r},t) = \int d^2 \mathbf{r}' K(\mathbf{r} - \mathbf{r}') \sum_{k'=1} ^N \frac{1}{N} \delta(\mathbf{r}' - \mathbf{r}_{k'} (t))\;.
\end{align}
Note that, in Eq.~\eqref{eq:v_qs}, $v(\varrho_\text{loc})$ is explicitly forbidden from becoming negative. This is because the quorum-sensing ABPs lack short-range repulsion and can form a high-density region where $\alpha n - \varrho_{\mathrm{loc}}\zeta < 0$. Meanwhile, the kernel $K(\mathbf{r})$ should be positive only within a finite range, which is chosen to be
\begin{align}
    K(\mathbf{r}) = \begin{cases}
        1/\pi &\text{if $|\mathbf{r}| < 1$,} \\
        0 &\text{otherwise.}
    \end{cases} 
\end{align}
This ensures that each particle decelerates according to the mean-field density in its neighborhood rather than via repulsive pairwise interactions. The model thereby suppresses the effects of microscopic fluctuations, which wash out the phase-separated patterns predicted by the hydrodynamic theory. Such mean-field interactions between particles is possible when they are capable of quorum sensing, hence the name quorum-sensing ABPs.

\begin{figure*}
	\includegraphics[width=0.99\textwidth]{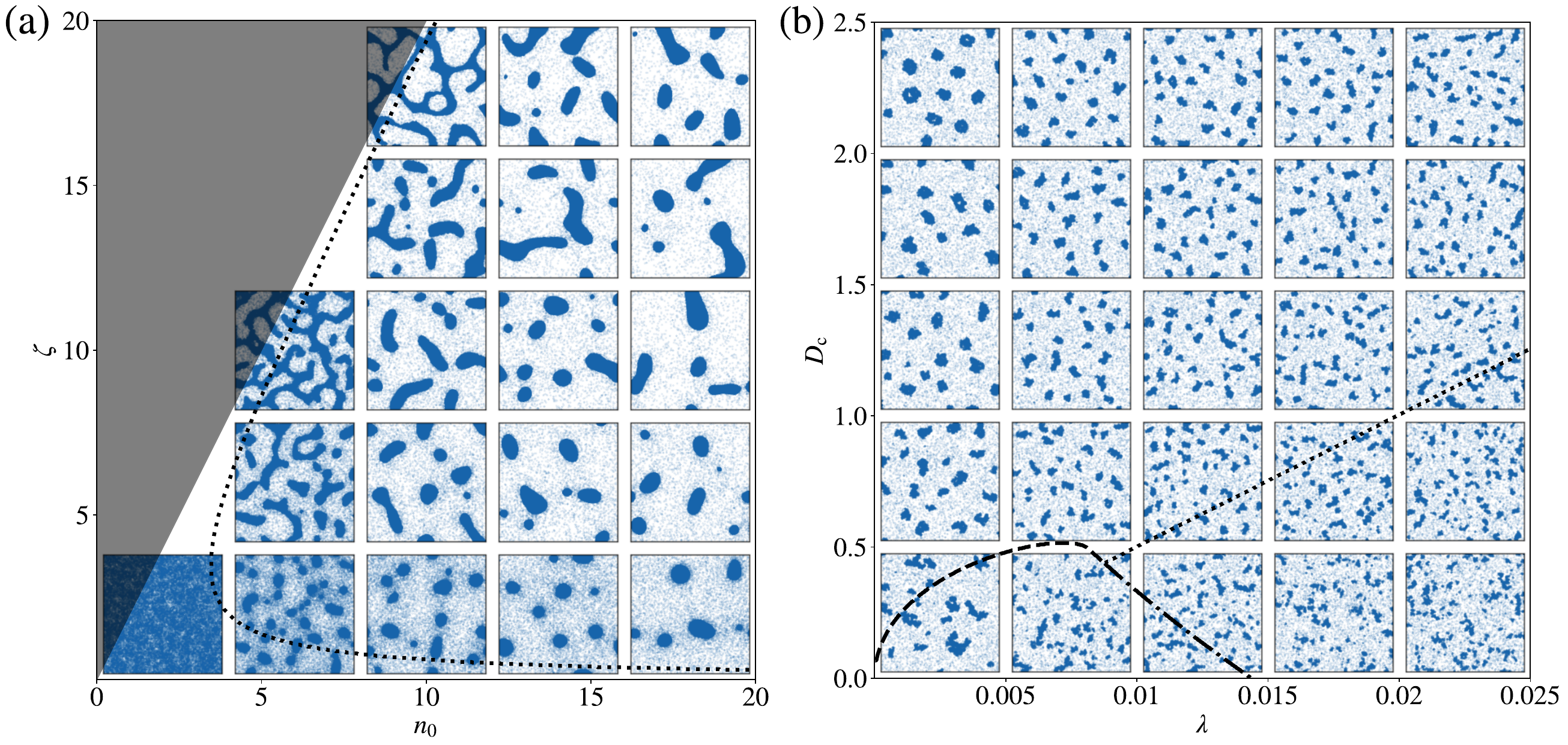}
    \caption{\label{fig:QS_phase_diagrams} \textbf{Snapshots and phase diagrams of the quorum-sensing active Brownian particles.} (a) The Basal Metabolic Regime as the global chemical concentration $n_0$ and the deceleration parameter $\zeta$ are varied for $\lambda = 0.5$, $D_\mathrm{c}=10$, $L=400$, $t=4\times10^2$ and (b) the Active Metabolic Regime as the chemical consumption rate $\lambda$ and the chemical diffusion coefficient $D_\mathrm{c}$ are varied for $n_0=20$, $\zeta=22$, $L=400$, $t=1.6\times10^3$. The global particle density is $\rho_0 = 1/2$ for both (a) and (b). The unphysical region with $v_\mathrm{eff} = \alpha n_0 - \rho_0 \zeta < 0$ is grayed out. Predictions of linear stability analysis for the phase boundaries (at $\kappa = 4.791667$) are indicated by the dotted (H--S boundary), the dash-dotted (H--O boundary), and the dashed (S--O boundary) lines. For comparison, snapshots of particle distribution (with darker regions indicating higher density) obtained by particle-based simulations are shown together. The central coordinates of each snapshot correspond to the parameters used in the simulation. Note that the snapshots in (b) have been cropped to the size of $200\times 200$ for better visibility. See Supplementary Movies 4A and 4B for animated versions of (a) and (b), respectively.
    }
\end{figure*}

Quorum-sensing ABPs have some advantages over mechanically interacting ABPs. First, the model's parameters are identical to those of the hydrodynamic model, making their quantitative comparison easier. The only hydrodynamic parameter not explicitly built into the model is $\kappa$, which is an independent parameter in the original hydrodynamic model defined by Eq.~\eqref{eq:hydro}. If we derive yet another hydrodynamic theory for the quorum-sensing ABPs by applying the coarse-graining scheme discussed in Appendix~A, the effective value of $\kappa$ for the quorum-sensing model can be expressed in terms of other parameters as follows:
\begin{align}
	\kappa = \frac{\rho_0\zeta(\alpha n_0 - \rho_0\zeta)}{8D_\mathrm{r}}-\frac{D}{D_\mathrm{r}}D_\mathrm{eff}\;.
\end{align}
For a ``fair'' comparison with the original hydrodynamic model, we can set $\kappa$ of the model equal to the value determined by the above equation. The second advantage is that, since there is no short-range repulsion, the simulation of quorum-sensing ABPs can work even with relatively large time steps without blowing up. Indeed, we use $\Delta t = 10^{-3}$ for the quorum-sensing ABPs, while $\Delta t = 10^{-5}$ has to be used for the mechanically interacting ABPs.

\begin{figure}[t]
\includegraphics[width=\columnwidth]{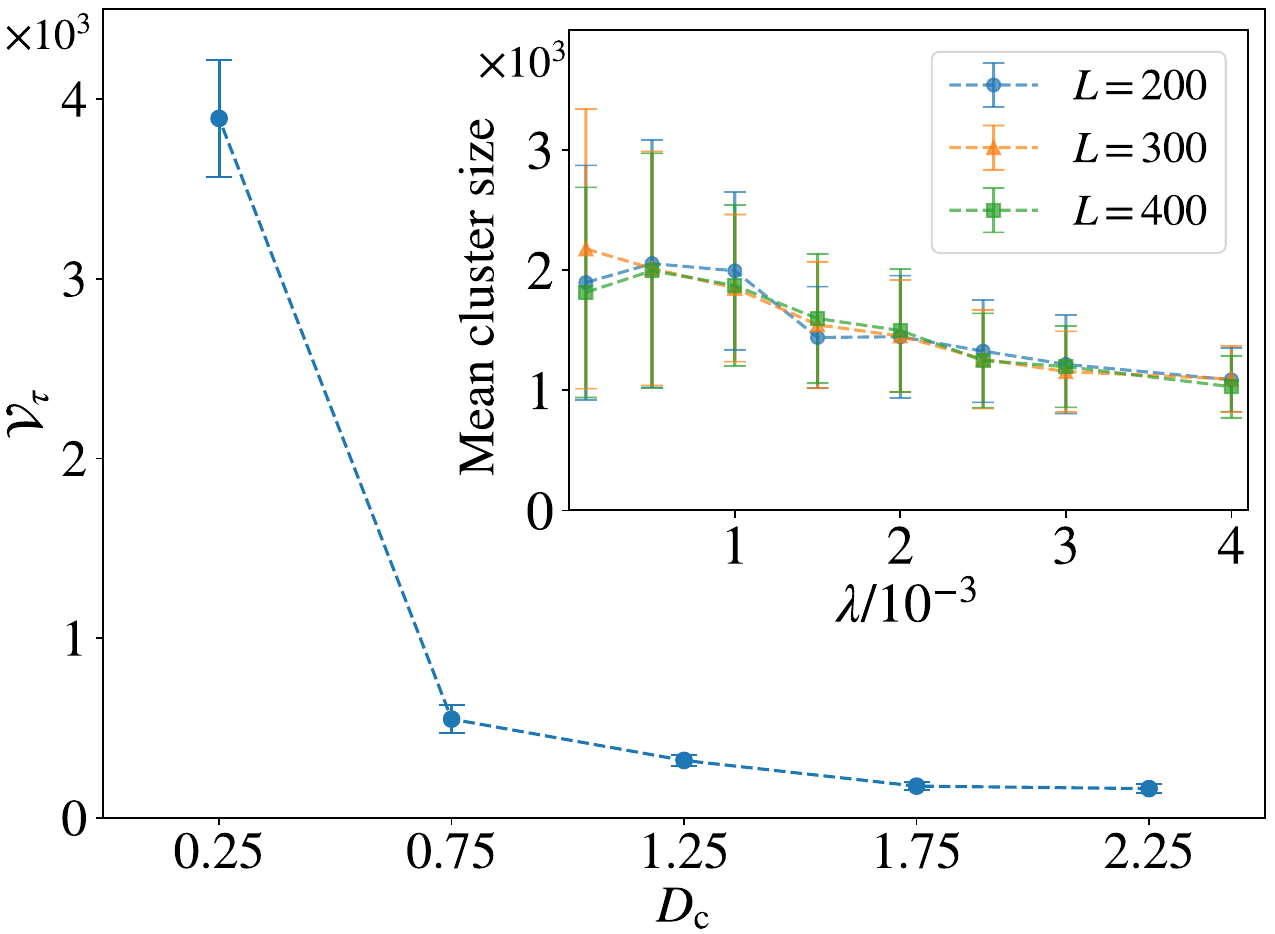}
    \caption{\label{fig:QS_TV_Vor} 
    \textbf{Properties of patterns formed by the quorum-sensing active Brownian particles in the Active Metabolic Regime.} The main figure shows how the time-variability $\mathcal{V}_\tau$ depends on the chemical diffusion constant $D_\mathrm{c}$ for $\tau=200$, $L_\mathrm{win}=35$, and $\lambda=0.0025$. Inset: The mean (symbols) and the standard deviation (error bars) of cluster sizes as functions of the chemical consumption rate $\lambda$ and the system size $L$ for $D_\mathrm{c}=2.5$ at $t=2\times 10^3$. For all plots, we use $n_0=20$ and $\zeta=22$.}
\end{figure}

Let us now examine how the behaviors of the quorum-sensing ABPs compare with the hydrodynamic model. For the BMR, as detailed in Methods, the boundary between the H and S phases is determined only by infinite-wavelength features of the system. The property, which is also reflected in the absence of $\kappa$ in Eq.~\eqref{eq:BMR_phase} for the phase boundary of the BMR, implies that the phase boundary of the quorum-sensing ABPs coincides exactly with that of the hydrodynamic theory. This is verified by Fig.~\ref{fig:QS_phase_diagrams}(a), which shows the snapshots of the simulations of the quorum-sensing ABPs at $t = 4\times 10^2$ (starting from uniform configurations at $t = 0$) together with the phase boundary predicted by Eq.~\eqref{eq:BMR_phase} (marked by the dotted line). As was the case for the hydrodynamic model (see Fig.~\ref{fig:AMR_time_evol}), the system tends to form more concentrated particle clusters for higher $n_0$, while sparse regions broaden at the expense of dense regions as $n_0$ is decreased. If the system is allowed to evolve even longer (to $t = 2 \times 10^3$), different clusters merge via the Ostwald process, forming large-scale patterns (see Fig.~\ref{fig:appendix_qs_BMR} and Appendix~\ref{sec:AppendixE}) also observed in the hydrodynamic model.

For the AMR, the short-wavelength features (which breaks the assumption of slow spatial variations used in the hydrodynamic theory) affect the phase boundaries, so the correspondence between the quorum-sensing ABPs and the hydrodynamic model cannot be exact. Still, Fig.~\ref{fig:QS_phase_diagrams}(b) and Supplementary Movie~4B suggest some qualitative agreements between the two. At fixed $D_\mathrm{c}$, the snapshots show that the particle clusters shrink with increasing $\lambda$. This is also confirmed by the $\lambda$-dependence of the mean cluster size at $D_\mathrm{c} = 2.5$ shown in the inset of Fig.~\ref{fig:QS_TV_Vor}, which also reveals that the mean cluster size does not change appreciably with the system size $L$. These results are qualitatively similar to how the S phase changes to the H phase in the hydrodynamic theory. The reason why the H phase is not observed within the simulated range of $\lambda$ may be an artifact generated by Eq.~\eqref{eq:v_qs}, which completely freezes small high-density regions satisfying $\alpha n - \varrho_{\mathrm{loc}}\zeta < 0$.

Meanwhile, when $\lambda$ is fixed at a small value, the snapshots in Fig.~\ref{fig:QS_phase_diagrams}(b) indicate that the particle clusters fluctuate more vigorously as $D_\mathrm{c}$ is reduced. As done in the case of the hydrodynamic model, we may use the mean vorticity to quantify this trend; however, since the calculation of vorticity requires coarse-graining the particle configuration that introduces extra numerical error, we resort to an alternative measure. For some $\tau > 0$, we define the time-variability
\begin{align}
    \mathcal{V}_\tau \equiv \int dt \int d^2 \mathbf{r}\, \big|\bar{n}(\mathbf{r}, t+\tau) - \bar{n}(\mathbf{r},t)\big|^2\;,
\end{align}
where $\bar{n}(\mathbf{r},t)$ is the average of the chemical concentration $n(\mathbf{r},t)$ over a square window of side length $L_\mathrm{win}$ centered at the position $\mathbf{r}$. By examining how large $\mathcal{V}_\tau$ can be for a suitable choice of $\tau$ and  $L_\mathrm{win}$, one can quantify the magnitude, time scale, and length scale of cluster fluctuations present in the system. At $\lambda = 0.0025$, choosing $\tau = 200$ and $L_\mathrm{win} = 35$, we observe a dramatic increase of $\mathcal{V}_\tau$ as $D_\mathrm{c}$ is lowered [see Fig.~\ref{fig:QS_TV_Vor}]. While $\mathcal{V}_\tau$ does not exhibit the dip near the S--O boundary shown by the mean vorticity in Fig.~\ref{fig:hydro_MSC_Vor}(f), this may be because the quantity is not as sensitive to the total length of cluster boundaries as the mean vorticity is. Moreover, the behaviors of $\mathcal{V}_\tau$ also suggest that the length scale of cluster patterns tends to decrease with increasing $\lambda$, as shown in Fig.~\ref{fig:appendix_qs_AMR} and Appendix~\ref{sec:AppendixE}. These support the existence of the S--O and the H--O boundaries predicted by the hydrodynamic theory.

We note that the ballistic bands and the radiating ripples generated by the hydrodynamic model are not observable for the quorum-sensing ABPs. It might be that such large-scale patterns are destroyed by the higher-order gradients neglected in the hydrodynamic theory or the microscopic fluctuations which are not entirely removed by introducing mean-field interactions. Still, our results thus far imply that particle-based models can exhibit both S and O phases predicted by the hydrodynamic theory.

\section*{Discussion}
\label{sec:summary}



We showed that chemokinesis combined with chemical consumption can profoundly affect the collective behaviors of active particles. In the BMR, where only a negligible portion of chemical consumption is related to particle motion, lower chemical concentration and weaker self-propulsion within particle clusters strengthen the tendency for MIPS. In the AMR, where chemical consumption is tightly coupled to particle motion, MIPS is suppressed due to higher chemical concentration and stronger self-propulsion within particle clusters. In this regime, MIPS occurs only as microphase separation. According to the hydrodynamic theory, this regime is divided further into the S phase, where the particle clusters are stable, and the O phase, where the clusters form, move, and vanish rapidly. While we are not aware of any experimental studies addressing the interplay of MIPS, chemokinesis, and chemical consumption, the predictions of our theory would be relevant to synthetic active particles with predominantly chemokinetic response to the fuel~\cite{Howse2007,Moran2021}.

We also note that the AMR's O phase resembles the dynamic clustering observed in the colloidal suspension of platinum--gold Janus particles~\cite{Theurkauff2012}, which was attributed to their chemotactic response to the fuel~\cite{Pohl2014dynamic,Pohl2015selfphoretic,Fadda2023interplay}. However, the O phase requires that the relaxation time scale of chemical concentration is comparable to or slower than that of particle density. To meet this criterion, the diffusion constant $D_\mathrm{c}$ of the chemical must be less than or comparable to the effective diffusion constant $D_\mathrm{eff}$ of the active particles. Experimental realizations (see \cite{BechingerRMP2016} and the references therein) of micron-size Janus particles in an aqueous solution of the fuel ({\em e.g.}, hydrogen peroxide) typically satisfy $D_\mathrm{eff} \approx 1~\mu\mathrm{m}^2 s^{-1}$ and $D_\mathrm{c} \approx 10^3~\mu\mathrm{m}^2 s^{-1}$, so the O phase is well beyond the regime of such empirical systems. To achieve $D_\mathrm{c} \lesssim D_\mathrm{eff}$, we must either reduce the size of the active particle to nanoscale (active particles as small as $50~\mathrm{nm}$ were achieved in \cite{Ma2015catalytic,*Ma2015enzyme}) or increase the self-propulsion velocity $v$ by a factor of $30$ (since $D_\mathrm{eff} \sim v^2$).

The dichotomy between the full phase separation of the BMR and the microphase separation of the AMR is reminiscent of the same two types of phase separations exhibited by the Active Model B+~\cite{Tjhung2018}. Whether there exists a common physical principle underlying such similar dichotomies is an interesting question. Moreover, it remains to be checked whether some higher-order terms or microscopic fluctuations neglected in the hydrodynamic theory destroy the large-scale oscillating patterns predicted for the AMR.

This study demonstrates the importance of chemical consumption mechanisms on the patterns of active matter. Future research may contain similar analyses for models beyond scalar active matter, including models with polar order, topological defects, etc. Moreover, chemical consumption may also modify various mechanical abilities of active matter, such as enhanced transport and current rectification. This may affect the performance of collective machines utilizing active matter and, therefore, should be investigated further.

\section*{Data availability}
The data that support the plots within this paper and other findings of this study are available from the corresponding author upon reasonable request.

\section*{Code availability}
The computer codes used to generate the results of this paper are available at \url{https://github.com/KwonEuiJoon/chemokinetic_active_matter}.

\bibliography{chemokinesis}

\section*{Acknowledgments}
This work was supported by the National Research Foundation of Korea (NRF) grants (RS-2021-017476, RS-2023-00218318, and RS-2023-00278985, RS-2024-00410147) funded by the Ministry of Science and ICT (MSIT) of the Korea government. We thank \'{E}tienne Fodor and Ignacio Pagonabarraga for helpful discussions.

\section*{Author contributions}
E.K. and Y.O. derived and analyzed the hydrodynamic equations and performed numerical simulations. Y.B. supervised the project. All authors participated in the writing of the manuscript.

\section*{Competing interests}
The authors declare no competing interests.


\appendix

\pagebreak
\widetext

\renewcommand{\arraystretch}{1.5}

\def\bibsection{\section*{Supplementary References}}


\setcounter{equation}{0}
\setcounter{figure}{0}
\setcounter{table}{0}
\makeatletter
\renewcommand{\theequation}{A\arabic{equation}}
\renewcommand{\thefigure}{A\arabic{figure}}
\renewcommand{\theHfigure}{A\arabic{figure}}
\renewcommand{\thetable}{A\arabic{table}}

\section{Derivation of hydrodynamic equations}\label{sec:appendixa}
Here we provide the details of how the hydrodynamic equations [Eq.~(\ref{eq:hydro}) in the main text] are derived from the particle-based models [Eqs.~(\ref{eq:eom}--\ref{eq:AMR_f}) in the main text]. The procedure closely follows the coarse-graining scheme given in~\cite{Bialke2013}. Denoting by $\varphi_k$ the orientational angle of the self-propulsion direction $\mathbf{e}_k = (\cos \varphi_k, \sin \varphi_k)$ of particle $k$, the time evolution of the normalized joint probability distribution $\psi_N (\{\mathbf{r}_k , \varphi _k \} ;t)$ is governed by 
\begin{align} \label{eq:varphiN}
    \partial_t \psi_N = \sum_{k=1} ^N \nabla_k \cdot [\mu (\nabla_k U) - \tilde{v} \mathbf{e}_k + D_0\nabla_k ]\psi_N  + D_\mathrm{r} \sum_{k=1} ^N \frac{\partial ^2 \psi_N}{\partial \varphi_k ^2}
    \;,
\end{align}
where $D_0 = \mu T$ and $U=\sum_{k\neq k'} V(|\mathbf{r}_k - \mathbf{r}_{k'}|)$. The first step is to integrate out the degrees of freedom associated with $N-1$ particles, which yields the projected density
\begin{align}
    \psi _1 (\mathbf{r}_1,\varphi_1 ; t) = \int d \mathbf{r}_2 \cdots d \mathbf{r}_N \int d\varphi _2 \cdots d\varphi _N N \psi _N
    \;
\end{align}
of a single particle. Since the particles are identical, from now on we drop the subscripts of $\mathbf{r}_1$ and $\varphi_1$, using $\mathbf{r}$ and $\varphi$ instead. Integrating Eq.~\eqref{eq:varphiN} side by side, we get
\begin{align} \label{eq:varphi1}
    \partial_t \psi _1 = -\nabla \cdot [\mu \mathbf{F} + \tilde{v} \,\mathbf{e}\, \psi _1 - D_0\nabla \psi_1] + D_\mathrm{r} \partial^2 _\varphi \psi_1\;,
\end{align}
with $\mathbf{e} \equiv (\cos\varphi, \sin\varphi)$ and the mean interaction force given by
\begin{align}
    \mathbf{F} (\mathbf{r},\varphi ;t) &\equiv \int d \mathbf{r}_2 \cdots d \mathbf{r}_N \int d\varphi _2 \cdots d\varphi _N (-\nabla U ) N \psi_N
    \nonumber \\ &= - \int d \mathbf{r}' \,V'(|\mathbf{r} - \mathbf{r}'|) \frac{\mathbf{r} - \mathbf{r}'}{|\mathbf{r} - \mathbf{r}'|} \psi _2 (\mathbf{r},\varphi,\mathbf{r}';t)
    \;.
\end{align}
Here, $\psi_2 (\mathbf{r},\varphi,\mathbf{r}';t)$ is the two-body density associated with the joint probability that there is a particle of orientation $\varphi$ at $\mathbf{r}$ while another is located at $\mathbf{r}'$.

Next, we aim to formally eliminate the dependence of Eq.~\eqref{eq:varphi1} on $\mathbf{F}$, replacing it with the dependence on the local particle density
\begin{align}
    \rho (\mathbf{r} ,t) \equiv \int_0 ^{2\pi} d\varphi\, \psi _1 (\mathbf{r},\varphi ,t)
    \;.
\end{align}
Then, the two-particle distribution can be reexpressed as
\begin{align}
    \psi_2 (\mathbf{r},\varphi,\mathbf{r}'\,;\, t) = \psi_1 (\mathbf{r},\varphi \,;\,t) \,\rho(\mathbf{r})\, g(\mathbf{r}' \,|\,\mathbf{r},\varphi \,;\, t )
    \;,
\end{align}
where $g(\mathbf{r}' \,|\,\mathbf{r},\varphi \,;\, t )$ is the conditional probability of finding another particle at $\mathbf{r}'$ given that there is a particle  at $\mathbf{r}$ with orientation $\varphi$. Further simplifications can be made by assuming that the system is effectively homogeneous, isotropic, and fast-relaxing when it comes to the two-particle statistics (despite the spatial and temporal variations of $\rho$ and $\mathbf{p}$). Under these assumptions, we can write
\begin{align}
    g(\mathbf{r}' \,|\,\mathbf{r},\varphi \,;\, t ) \simeq g(|\mathbf{r} - \mathbf{r}'|,\theta)\;,
\end{align}
so that only the distance $|\mathbf{r} - \mathbf{r}'|$ and the directional angle $\theta$ matter. Then, we get $\mathbf{e} \cdot \mu\mathbf{F} =-\rho \zeta \psi_1$ with
\begin{align} \label{eq:zeta_def}
    \zeta \equiv -\mu \int_0 ^\infty dr \,r \,V'(r) \int_0 ^{2\pi} d\theta \,\cos \theta \,g(r,\theta)
    \;.
\end{align}
We may assume that $\phi_1$ is slow-varying in space, which allows us to decompose $\mathbf{F}$ linearly in terms of the orientation $\mathbf{e}$ and the density gradient $\nabla \psi _1$. This yields
\begin{align}
    \mathbf{F} = (\mathbf{e} \cdot \mathbf{F} ) \mathbf{e} + (D_0-D) \nabla \psi_1 + \mathcal{O}(|\nabla \psi_1|^2 )
    \;,
\end{align}
where $D \equiv D_0 - \mathbf{F} \cdot [\nabla \psi_1 - \mathbf{e} (\mathbf{e} \cdot \nabla \psi_1 )]/|\nabla \psi _1 |$. Using the relation in Eq.~\eqref{eq:varphi1}, the equation of motion for $\psi_1$ is simplified to
\begin{align} \label{eq:varphi1_noF}
    \partial_t \psi_1 = -\nabla \cdot [v \mathbf{e} -D\nabla]\psi_1 + D_\mathrm{r} \partial^2 _\varphi \psi _1
    \;,
\end{align}
where $v \equiv \tilde{v} - \rho \zeta$. Thus, we have formally replaced the effects of the interaction force $\mathbf{F}$ with the deceleration effect governed by $-\rho\zeta$ and the renormalized diffusion coefficient $D$. While $D$ defined above is in principle not uniform in space, we resort to the effective homogeneity of the system and treat $D$ as a constant in the hydrodynamic equations.

While Eq.~\eqref{eq:varphi1_noF} has no explicit dependence on $\mathbf{F}$, it is still dependent on the particle orientation $\mathbf{e}$. To eliminate the dependence, we define the polarization field
\begin{align}
    \mathbf{p} (\mathbf{r},t) \equiv \int_0 ^{2\pi} d \varphi \, \mathbf{e} \, \psi_1 (\mathbf{r},\varphi ,t)\;.
\end{align}
Then, integrating Eq.~\eqref{eq:varphi1_noF} side by side with respect to $\varphi$ before and after multiplying $\mathbf{e}$, we obtain
\begin{align}
    \partial_t \rho (\mathbf{r} ,t) &= -\nabla \cdot [ v \mathbf{p} (\mathbf{r},t) - D \nabla \rho (\mathbf{r},t) ] \label{eq:rho}
    \\ \partial_t \mathbf{p} (\mathbf{r}, t) &= - \frac{1}{2} \nabla (v \rho)  + D \nabla^2 \mathbf{p} - D_\mathrm{r} \mathbf{p}\;,
\end{align}
where we neglected the nematic contributions assuming that they quickly relax to zero. We further assume that $\mathbf{p}(\mathbf{r},t)$ also quickly relaxes to the steady state determined by $\rho(\mathbf{r},t)$, with slow spatial variations. Then, at the leading order, we can use the approximation $\mathbf{p}(\mathbf{r},t) \simeq -\nabla(v\rho) / (2D_\mathrm{r})$. Plugging this expression into Eq.~\eqref{eq:rho} yields
\begin{align}
    \partial_t \rho = \nabla \cdot \left[D \nabla \rho + \frac{v}{2D_\mathrm{r}} \nabla (v\rho) \right]\;.
\end{align}ion
To obtain Eq.~(\ref{eq:hydro}) in the main text, we need to introduce an additional term $\kappa \nabla^2 (\nabla^2 \rho)$ on the right-hand side of the above equation, which is the next lowest-order term allowed by the symmetry of the system and phenomenologically necessary to account for the interface energy between dense and dilute regions. Microscopically, the term would naturally arise from higher-order gradients of the fields that were neglected in this derivation.

Now, we derive the hydrodynamic equation for the chemical. Assuming that sufficiently many active Brownian particles (ABPs) interact with the chemical at every location, the chemical concentration $n(\mathbf{r},t)$ can be regarded as a deterministic field, with the consumption function $f$ in Eq.~(\ref{eq:eom_chem}) in the main text replaced with its average over the particle distribution $\psi_N$, which we denote by $\langle f \rangle$. For the Basal Metabolic Regime (BMR), it is straightforward to see that
\begin{align} \label{eq:BMR_f_avg}
    \langle f(n,\{\mathbf{r}_k,\dot{\mathbf{r}}_k \}) \rangle = n(\mathbf{r}) \sum_{k'=1} ^N \langle \delta(\mathbf{r}-\mathbf{r}_{k'})\rangle = n(\mathbf{r})\rho(\mathbf{r})
    \;.
\end{align}
For the Active Metabolic Regime (AMR), we rewrite $\dot{\mathbf{r}}_k$ in terms of the three terms on the right-hand side of Eq.
~(\ref{eq:eom}) of the main text and calculate the contribution of each term to $\langle f \rangle$ separately. The first term contributes
\begin{align}
    \sum_{k'=1} ^N \left\langle  \tilde{v}(n(\mathbf{r}_{k'})))\delta(\mathbf{r}-\mathbf{r}_{k'}) \right\rangle =  \tilde{v}(n(\mathbf{r}))\rho(\mathbf{r})
    \;.
\end{align}
Meanwhile, the second term contributes
\begin{align}
    &\sum_{k \neq k'} \mu \left\langle \mathbf{e}_{k} \cdot \nabla_k V(|\mathbf{r}_k - \mathbf{r}_{k'}|) \delta(\mathbf{r}-\mathbf{r}_k)\right\rangle \nonumber
    \\ &= \sum_k \int d\mathbf{r}_1 \cdots d\mathbf{r}_N d\varphi_1 \cdots d\varphi_N\, \mathbf{e}_k \cdot \mu \nabla_k U\delta(\mathbf{r}-\mathbf{r}_k) \psi _N \nonumber \\ 
    &= -\frac{1}{N}\sum_k  \int d\mathbf{r}_k d\varphi_k\, \mathbf{e}_k \cdot \mu \mathbf{F}(\mathbf{r}_k,\varphi_k ;t) \delta(\mathbf{r} - \mathbf{r} _k )
    \nonumber \\ &= \frac{1}{N} \sum_k \int d\mathbf{r}_k d\varphi_k \,\rho(\mathbf{r}_k) \zeta \psi_1 (\mathbf{r}_k, \varphi_k)\delta(\mathbf{r}-\mathbf{r}_k) 
    \nonumber \\ &= \rho(\mathbf{r})^2 \zeta
    \;,
\end{align}
where we used the relation above Eq.~\eqref{eq:zeta_def} to obtain the third equality. Finally, the third term has zero contribution as the white noise $\bm\xi_k$ averages to zero. Combining the above results, the AMR satisfies
\begin{align} \label{eq:AMR_f_avg}
    \langle f(n,\{\mathbf{r}_k,\dot{\mathbf{r}}_k\})\rangle = n \rho (\tilde{v} - \rho \zeta) = n\rho v(\rho ,n)
    \;.
\end{align}
Using Eqs.~(\ref{eq:BMR_f}) and (\ref{eq:AMR_f}) to approximate Eq.~(\ref{eq:eom_chem}) in the main text, we obtain the hydrodynamic equation for the chemical concentration $n(\mathbf{r},t)$ shown in Eq.~(\ref{eq:hydro}).

\newpage
\section{Details of linear stability analysis}
\label{sec:AppendixB}
Here we provide details of linear stability analysis (LSA) for the prediction of MIPS.
Suppose that the density $\rho$ and the concentration $n$ are weakly perturbed around their homogeneous values $\rho_0$ and $n_0$, respectively:
\begin{align}
\rho = \rho_0 + \delta\rho ,\,\,n=n_0 + \delta n
\;.
\end{align}
Using these in Eq.~(\ref{eq:v_rhon}) of the main text, the time evolution of $\delta\rho$ is expanded to linear order as
\begin{align} \label{eq:drho}
    \delta \dot{\rho} = D_{\text{eff}} \nabla ^2 \delta \rho - \kappa \nabla^2 (\nabla ^2 \delta \rho) + \frac{\alpha \rho_0 (\alpha n_0 - \rho_0 \zeta)}{2D_\mathrm{r}} \nabla^2 \delta n\;,
\end{align}
where
\begin{align}
    D_{\text{eff}} \equiv D + \frac{(\alpha n_0 - \rho_0 \zeta )(\alpha n_0 - 2\rho_0 \zeta)}{2D_\mathrm{r}}
\end{align}
is the effective diffusion coefficient governing the collective relaxation of the density profile. Meanwhile, $\delta n$ evolves in the linear regime according to
\begin{align} \label{eq:dn_BMR}
    \delta \dot{n} = D_\mathrm{c} \nabla^2 \delta n - \lambda n_0 \,\delta \rho - \lambda \rho_0 \,\delta n 
\end{align}
for the BMR and
\begin{align} \label{eq:dn_AMR}
    \delta \dot{n} = D_\mathrm{c} \nabla^2 \delta n - \lambda n_0 (\alpha n_0 - 2\rho_0 \zeta)\, \delta \rho  -\lambda \rho_0 (2\alpha n_0 - \rho_0 \zeta)\, \delta n 
\end{align}
for the AMR.

Now, taking the Fourier components $\delta \rho = \delta \rho _0  e^{i\mathbf{q}\cdot \mathbf{r} + \omega t}$ and $\delta n = \delta n _0  e^{i\mathbf{q}\cdot \mathbf{r} + \omega t}$, Eqs.~\eqref{eq:drho}, \eqref{eq:dn_BMR}, and \eqref{eq:dn_AMR} can be rewritten as an eigenvalue equation
\begin{align}
      \begin{pmatrix} \mathcal{A} & \mathcal{B} \\ \mathcal{C} & \mathcal{D} \end{pmatrix} \begin{pmatrix} \delta \rho_0 \\ \delta n_0 \end{pmatrix} = \omega \begin{pmatrix} \delta \rho_0 \\ \delta n_0 \end{pmatrix}\;,
\end{align}
where
\begin{align} \label{eq:ABCD}
\mathcal{A} &= - \kappa q^4 - D_{\text{eff}}\,q^2\;,
\nonumber\\ \mathcal{B} &= - \left[\frac{\alpha \rho_0 (\alpha n_0 - \rho_0 \zeta)}{2D_\mathrm{r}} \right]q^2\;,
\nonumber\\ \mathcal{C} &= \begin{cases} -\lambda n_0 & \text{(BMR)} \\ -\lambda n_0 (\alpha n_0 - 2\rho_0 \zeta)  & \text{(AMR)} \end{cases}\;, \nonumber
\\ \mathcal{D} &= \begin{cases} -D_\mathrm{c} q^2 - \lambda \rho_0 & \text{(BMR)} \\ -D_\mathrm{c} q^2 - \lambda \rho_0(2\alpha n_0 - \rho_0 \zeta )  & \text{(AMR)} \end{cases}
\;.
\end{align}
Solving the characteristic equation
\begin{align} \label{eq:omega_char}
    \omega ^2 - (\mathcal{A}+\mathcal{D})\omega + \mathcal{A}\mathcal{D} - \mathcal{B}\mathcal{C} = 0
    \;,
\end{align}
the eigenvalues are obtained as
\begin{align} \label{eq:omega_pm}
\omega_{\pm} = \frac{1}{2} \left[(\mathcal{A}+\mathcal{D}) \pm \sqrt{(\mathcal{A}-\mathcal{D})^2 + 4\mathcal{B}\mathcal{C}} \right]
\;.
\end{align}
Among the two eigenvalues, the one with the larger real part always dominates the long-time behavior. If $(\mathcal{A}-\mathcal{D})^2 + 4\mathcal{B}\mathcal{C}>0$, then $\omega_+$ is the dominant one. If $(\mathcal{A}-\mathcal{D})^2 + 4\mathcal{B}\mathcal{C} \le 0$, then both $\omega_+$ and $\omega_-$ are equally dominant.

\subsubsection{Case of the BMR}
In this case, $\mathcal{B}\mathcal{C}$ is always positive. Then Eq.~\eqref{eq:omega_pm} ensures that both eigenvalues are real, with the boundary between the H and the S phases determined by when $\omega_+$ changes its sign. In the S phase, near the phase boundary, we must have $\omega_+ > 0$ and $\omega_- < 0$. By examining the quadratic left-hand side of Eq.~\eqref{eq:omega_char}, one can see that this is equivalent to the condition $\mathcal{A}\mathcal{D}-\mathcal{B}\mathcal{C} < 0$. By simple algebra, we obtain $\mathcal{A}\mathcal{D}-\mathcal{B}\mathcal{C} = q^2h(q^2)$, where
\begin{align}
 h(q^2) = \kappa D_\mathrm{c} q^4 +(D_\mathrm{eff}D_\mathrm{c}+\kappa\lambda \rho_0)q^2 +\lambda\rho_0D_\mathrm{eff} - \frac{\lambda\rho_0 }{2D_\mathrm{r}} \alpha n_0(\alpha n_0 - \rho_0 \zeta)
\end{align}
is a quadratic function of $q^2$. For $h(q^2) < 0$ to be satisfied for some range of $q^2$, there are two possible cases. First, if $h(0) < 0$, there exists an interval $q^2 \in [0,\,q^2 _*)$ in which $h(q^2) < 0$ holds. This condition is equivalent to
\begin{align} \label{eq:Deff_boundary}
    D_\text{eff} < \frac{\alpha n_0}{2D_\mathrm{r}}(\alpha n_0 - \rho_0 \zeta).
\end{align}
Second, if $h(0) > 0$, then $h'(0) < 0$ must hold for $h(q^2)$ to be negative for some interval $q^2 \in [q^2_-,\,q^2_+]$. In this case,
\begin{align}
    D_\text{eff} > \frac{\alpha n_0}{2D_\mathrm{r}}(\alpha n_0 - \rho_0 \zeta)
\end{align}
and
\begin{align}
    D_\text{eff} < -\frac{\kappa\lambda\rho_0}{D_\text{c}}
\end{align}
must be satisfied simultaneously, which is impossible. Thus, only Eq.~\eqref{eq:Deff_boundary} determines the S phase, which can also be written as
\begin{align}
    D < \frac{\rho_0 \zeta (\alpha n_0 - \rho_0 \zeta)}{D_\mathrm{r}}.
\end{align}
The phase boundary obtained in Eq.~(\ref{eq:eig_eq}) follows directly from this result. Notably, the discussion thus far shows that the boundary of the H phase in the BMR is determined only by infinite-wavelength fluctuations ($q = 0$), reflecting that the BMR is capable of large-scale phase separation even in the linear regime.

Meanwhile, in the absence of particle--chemical coupling ($\lambda = 0$), $h(q^2) < 0$ holds for some range of $q^2$ if and only if $h'(0) = D_\text{eff}D_\text{c} < 0$. This implies that the vanilla model exhibits MIPS for $D_\text{eff} < 0$ (which is clearly more stringent than Eq.~(\ref{eq:eig_eq}) in the main text for the BMR), from which Eq.~(\ref{eq:BMR_phase}) in the main text is derived.

\subsubsection{Case of the AMR}

In this case, $\mathcal{B}\mathcal{C}$ may be negative, allowing for nonvanishing imaginary parts of the eigenvalues $\omega_\pm$. Indeed, it turns out that the AMR exhibits all the three different phases (H, S, and O). In the following, we elaborate on how to determine the boundaries of those phases.

Let us first determine the boundaries of the H phase, which requires that the real part of $\omega_\pm$ is negative for all $q^2$. Due to the quadratic nature of Eq.~\eqref{eq:omega_char}, the condition $\mathrm{Re}~\omega_\pm < 0$ is fulfilled when both $\mathcal{A}\mathcal{D} - \mathcal{B}\mathcal{C} > 0$ and $\mathcal{A}+\mathcal{D} < 0$ are true.

We start with the first inequality. By some algebra, we obtain $\mathcal{A}\mathcal{D} - \mathcal{B}\mathcal{C} = q^2 h(q^2)$, where
\begin{align} \label{eq:h_AMR}
h(q^2) &= \kappa D_\mathrm{c} q^4+[D_\mathrm{eff}D_\mathrm{c} + \kappa\lambda\rho_0(2\alpha n_0-\rho_0\zeta)]q^2 + \lambda\rho_0 D_\mathrm{eff}(2\alpha n_0- \rho_0\zeta)\nonumber\\
&\quad -\frac{\lambda\rho_0}{2D_\mathrm{r}}\alpha n_0 (\alpha n_0-\rho_0\zeta)(\alpha n_0-2\rho_0\zeta)
\end{align}
is a quadratic function of $q^2$. If the coefficient of $q^2$ is positive, the minimum of $h(q^2)$ is located at $q^2 = 0$. Then, $\mathcal{A}\mathcal{D} - \mathcal{B}\mathcal{C} > 0$ is equivalent to $h(0) > 0$. These imply
\begin{align} \label{eq:Deff_AMR_boundary1}
	D_\mathrm{eff} > \max\bigg[-\frac{\kappa\lambda\rho_0}{D_\mathrm{c}}(2\alpha n_0-\rho_0\zeta), \frac{\alpha n_0 (\alpha n_0-\rho_0\zeta)(\alpha n_0-2\rho_0\zeta)}{2D_\mathrm{r}(2\alpha n_0- \rho_0\zeta)}\bigg]\;.
\end{align}
On the other hand, if the coefficient of $q^2$ is negative, the minimum of $h(q^2)$ is located at the axis of symmetry $q^2 = -[D_\mathrm{eff}D_\mathrm{c} + \kappa\lambda\rho_0(2\alpha n_0-\rho_0\zeta)]/(2\kappa D_\mathrm{c})$. Thus, $\mathcal{A}\mathcal{D} - \mathcal{B}\mathcal{C} > 0$ requires
\begin{align} \label{eq:Deff_AMR_boundary2}
    &\lambda \rho_0 D_\mathrm{eff} (2\alpha n_0 - \rho_0 \zeta ) 
    - \frac{\lambda\rho_0}{2D_\mathrm{r}}\alpha n_0 (\alpha n_0-\rho_0\zeta)(\alpha n_0-2\rho_0\zeta)
    \nonumber\\ 
    &- \frac{1}{4\kappa D_\mathrm{c}} \left[D_\mathrm{eff}D_\mathrm{c} + \kappa\lambda\rho_0(2\alpha n_0-\rho_0\zeta) \right]^2 > 0
    \;,
\end{align}
which is quadratic in $D_\mathrm{eff}$ and can be solved analytically.

Next, we check $\mathcal{A} + \mathcal{D} < 0$, which can be written as a quadratic inequality
\begin{align}
    \kappa q^4 + (D_{\text{eff}} + D_\mathrm{c} )q^2 + \lambda\rho_0 (2\alpha n_0 - \rho_0 \zeta ) > 0
    \;.
\end{align}
The inequality is always satisfied when $D_{\text{eff}} + D_\mathrm{c} > 0$, as all the other coefficients are positive. When $D_{\text{eff}} + D_\mathrm{c} \le 0$, the minimum value of the quadratic function must be positive, that is,
\begin{align}
    \lambda \rho_0 (2\alpha n_0 - \rho_0 \zeta ) - \frac{(D_{\text{eff}} + D_\mathrm{c} )^2}{4\kappa} > 0 \;.
\end{align}
Combining all these results, $\mathcal{A} + \mathcal{D} < 0$ is equivalent to
\begin{align} \label{eq:Deff_AMR_boundary3}
D_\text{eff} > -D_\text{c} - 2\sqrt{\kappa\lambda\rho_0(2\alpha n_0 - \rho_0\zeta)}\;.
\end{align}

Based on the above, one can fully determine the boundaries of the H phase. Here we refrain from giving the explicit expressions for the corresponding values of $D_\mathrm{eff}$, which is a rather tedious task. Instead, we note that setting $\lambda > 0$ broadens the ranges of $D_\mathrm{eff}$ associated with the H phase. This can be seen through the following argument. First, we observe that Eqs.~\eqref{eq:Deff_AMR_boundary1} and \eqref{eq:Deff_AMR_boundary2} imply
\begin{align}
	D_\text{eff} > \frac{\alpha n_0 (\alpha n_0-\rho_0\zeta)(\alpha n_0-2\rho_0\zeta)}{2D_\mathrm{r}(2\alpha n_0- \rho_0\zeta)}\;.
\end{align}
Using this inequality in Eq.~\eqref{eq:h_AMR}, we obtain
\begin{align}
	h(q^2) \ge \kappa D_\mathrm{c} q^4 + D_\mathrm{eff}D_\mathrm{c} q^2\;,
\end{align}
with the equality holding if and only if $\lambda = 0$. This means that $\mathcal{A}\mathcal{D} - \mathcal{B}\mathcal{C} > 0$ becomes weaker for $\lambda > 0$ than for $\lambda = 0$. Similarly, Eq.~\eqref{eq:Deff_AMR_boundary3} shows that $\mathcal{A} + \mathcal{D} < 0$ becomes weaker for $\lambda > 0$ than for $\lambda = 0$. Since both inequalities are weakened by positive $\lambda$, we conclude that the AMR suppresses the MIPS and widens the H phase.

\begin{table}   
   \small
   \centering
   \begin{tabular}{c|c|c}
   \toprule\toprule
   & $y<4x(1-2x)$ & $y>4x(1-2x)$  \\ [0.5ex]
   \hline 
   \midrule 
   $x>1/3$ & $q_l^2 < \bm{q_- ^2} <q_r ^2 <\bm{q_+ ^2}$ & $q_l ^2 <\bm{q_- ^2} <\bm{q_+ ^2} < q_r ^2 $ \\ [0.5ex]
   $1/5 < x < 1/3$ & $q_l ^2 < \bm{q_- ^2} < q_r ^2 < \bm{q_+ ^2}$ & $q_l ^2 <q_r ^2 < \bm{q_- ^2} < \bm{q_+ ^2}$ \\ [0.5ex]
   $x<1/5$ & $q_l ^2 < \bm{q_- ^2} < q_r ^2 < \bm{q_+ ^2}$ & $\bm{q_- ^2} < q_l ^2 < q_r ^2 < \bm{q_+ ^2}$ \\

   \bottomrule
   \end{tabular}
   \caption{\label{tab:ordering} Ordering of $q_{\text{l,r}}^2$ and $q_{\pm}^2$. $q_{\pm}^2$ are bold-faced to emphasize the changes when transitioning from $y < 4x(1-2x)$ to $y > 4x(1-2x)$.} 
\end{table}

Now, we turn to the boundaries of the O phase, which corresponds to the case where $\omega_\pm$ have nonzero imaginary parts (requiring $(\mathcal{A}-\mathcal{D})^2+4\mathcal{B}\mathcal{C} < 0$) simultaneously with positive real parts (requiring $\mathcal{A} + \mathcal{D} > 0$).

Let us first check the condition for nonzero imaginary parts. Defining $f(q^2) \equiv (\mathcal{A}-\mathcal{D})^2$ and $g(q^2) \equiv -4\mathcal{B}\mathcal{C}$, the relevant inequality is given by $f(q^2) < g(q^2)$. We note that $f(q^2)$ is the square of a quadratic function with one positive root, which we denote by
\begin{align}
    q_c^2 \equiv -\frac{D_{\text{eff}} - D_\mathrm{c}}{2\kappa} + \frac{\sqrt{(D_{\text{eff}}-D_\mathrm{c})^2 + 4\kappa \lambda \rho_0 (2\alpha n_0 - \rho_0 \zeta) }}{2\kappa}
    \;.
\end{align}
Hence, $f(q^2)$ has a double root at $q^2 = q_c^2$.

\begin{figure}
\includegraphics[width=\columnwidth]
{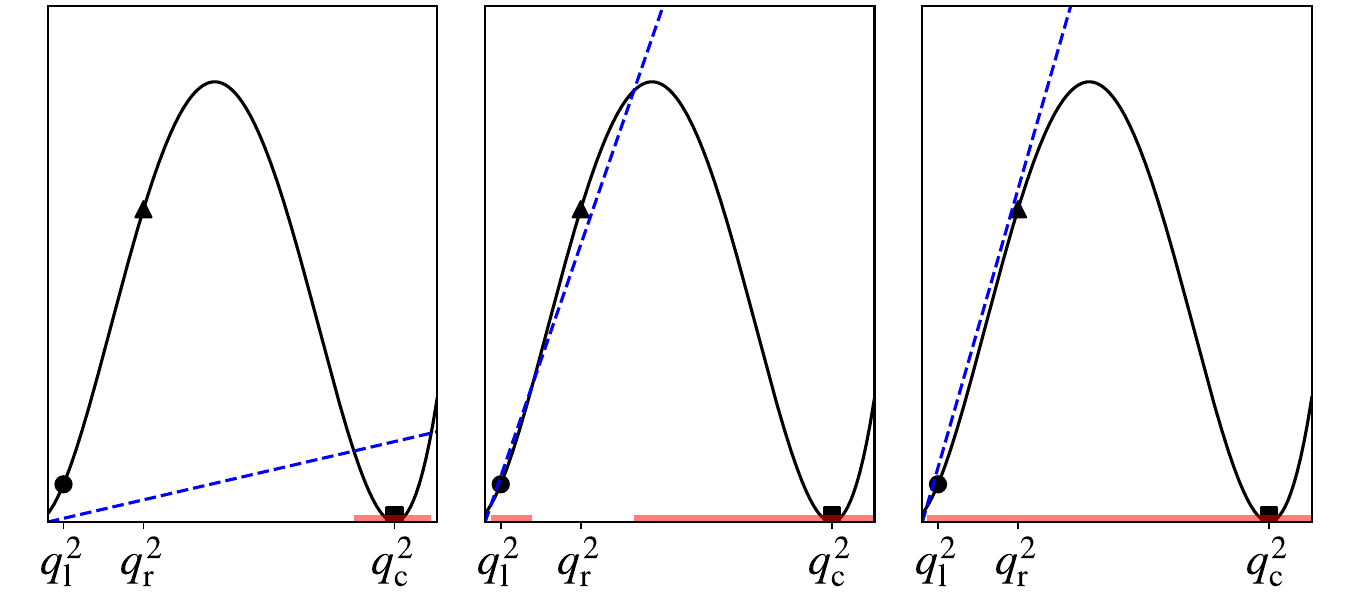}
    \caption{
    \textbf{Wave number intervals leading to oscillating fluctuations.} The black solid line represents the quadratic function $f(q^2)$, and the blue dashed line corresponds to the linear function $g(q^2)$. The points of tangency at $q_\mathrm{l} ^2$, $q_\mathrm{r} ^2$, and $q_\mathrm{c} ^2$ are marked by the circle, the triangle, and the square, respectively. The intervals of $q^2$ with $f(q^2) < g(q^2)$, which produce oscillating fluctuations, are indicated by thick red horizontal segments.}
    \label{fig:appendix_AMR_LSA}
\end{figure}

Meanwhile, $g(q^2)$ is a linear function of $q^2$ crossing the origin. Thus, to examine when $f(q^2) < g(q^2)$ is satisfied, it is helpful to check how many lines crossing the origin can be tangent to $f(q^2)$. The point of tangency can be found by solving
\begin{align}
    \frac{f(q^2)}{q^2} = \frac{d}{d(q^2)} f(q^2)
    \;,
\end{align}
which is equivalent to
\begin{align}
    3\kappa q^4 + (D_{\text{eff}} - D_\text{c} ) q^2 + \lambda \rho_0 (2\alpha n_0 - \rho_0 \zeta) = 0
    \;.
\end{align}
This equation has real roots if its discriminant satisfies
\begin{align} \label{eq:lr_exist}
    (D_\text{eff} - D_\text{c})^2 - 12 \kappa \lambda \rho_0 (2\alpha n_0 - \rho_0 \zeta) \ge 0
    \;.
\end{align}

We first consider the case where the above condition is satisfied. Then, the points of tangency are obtained as
\begin{align}
    q_{\text{l},{\text{r}}}^2 = -\frac{D_{\text{eff}}-D_\mathrm{c}}{6\kappa}
    \pm \frac{\sqrt{(D_{\text{eff}} - D_\mathrm{c})^2 - 12 \kappa \lambda \rho_0(2\alpha n_0 - \rho _0 \zeta )}}{6\kappa}
    \;,
\end{align}
where the plus (minus) sign corresponds to $q_\mathrm{r}$ ($q_\mathrm{l})$. We note that both $q_\mathrm{l}^2$ and $q_\mathrm{r}^2$ are positive if they have real values. As illustrated in Fig.~\ref{fig:appendix_AMR_LSA}, depending on the slope of $g(q^2)$ (indicated by the dashed line), the intervals of $q^2$ satisfying the condition $f(q^2) < g(q^2)$ (indicated by the thick horizontal lines) vary. When the slope of $g(q^2)$ is less than that of the tangent line crossing $(q_\mathrm{l}^2,f(q_\mathrm{l}^2))$ (indicated by the circle), there exists only a single interval satisfying $f(q^2) < g(q^2)$. If the slope of $g(q^2)$ is greater than that of the tangent line crossing $(q_\mathrm{r}^2,f(q_\mathrm{r}^2))$, the inequality is satisfied everywhere. In the intermediate regime, there are two separate intervals satisfying the inequality, one around $q_\mathrm{l}^2$ and the other around $q_\mathrm{c}^2$. As discussed earlier, $\omega_\pm$ have nonzero imaginary parts in these intervals.

The system exhibits the O phase if and only if these intervals of $q^2$ overlap with those where $\omega_\pm$ have positive real parts, which requires $\mathcal{A} + \mathcal{D} > 0$. This inequality is satisfied in the interval $(q_{-}^2, q_{+}^2)$, where
\begin{align}
    q_{\pm}^2 = -\frac{D_{\text{eff}} + D_\mathrm{c}}{2\kappa} 
     \pm \frac{\sqrt{(D_{\text{eff}} + D_\mathrm{c})^2 - 4\kappa \lambda \rho_0 (2\alpha n_0 - \rho_0 \zeta)}}{2\kappa}
    \;.
\end{align}
To investigate when the intervals overlap, we examine the condition for $q_{\text{l,r}}^2 = q_{\pm}^2$, where either of the two values on the left-hand side can be equal to either of the two on the other side. For convenience, we define $x \equiv -D_\mathrm{c} / D_{\text{eff}}$ and $y \equiv 4\kappa \lambda \rho_0 (2\alpha n_0 - \rho_0 \zeta) / D_{\text{eff}}^2$. After some algebra, one can show that $q_{\text{l,r}}^2 = q_{\pm}^2$ is satisfied when 
\begin{align}
    y = 4x(1 - 2x)
    \;.
\end{align}
This equation is useful for checking when the ordering of $q_{\text{l,r}}^2$ and $q_{\pm}^2$ changes. See Table~\ref{tab:ordering} for all possible orderings. For most cases, we have $q_\mathrm{l}^2 < q_-^2 < q_+^2$. Then the interval $(q_-^2,q_+^2)$ overlaps with nonzero imaginary parts of $\omega_\pm$ if and only if $\Delta(q^2) \equiv f(q^2) - g(q^2)$ becomes negative at $q^2 = q_{-}^2$ or $q_{+}^2$. In contrast, for $x < 1/5$ and $y > 4x(1 - 2x)$, the ordering changes to $q_-^2 < q_\mathrm{l}^2 < q_+^2$. It is then possible that $(q_- ^2 , q_+ ^2)$ fully contains an interval around $q_\mathrm{l}^2$ with nonzero imaginary parts of $\omega_\pm$, \textit{e.g.}, the left one of the two intervals indicated by thick horizontal segments in the middle panel of Fig.~\ref{fig:appendix_AMR_LSA}. For this particular case, the O phase requires that at least one among $\Delta(q_-^2)$, $\Delta(q_\mathrm{l}^2)$, and $\Delta(q_+^2)$ must be negative.



Finally, if Eq.~\eqref{eq:lr_exist} is not satisfied, then $f(q^2) < g(q^2)$ can hold only across a single continuous interval. Then, the O phase only requires that $\Delta(q_+ ^2)$ or $\Delta(q_- ^2)$ be negative. By combining all the conditions discussed so far, the boundaries of the O phase can be fully determined.

\newpage
\section{Effects of upper bound on self-propulsion}
\label{sec:AppendixC}
\begin{figure}
\includegraphics[width=0.6\columnwidth]{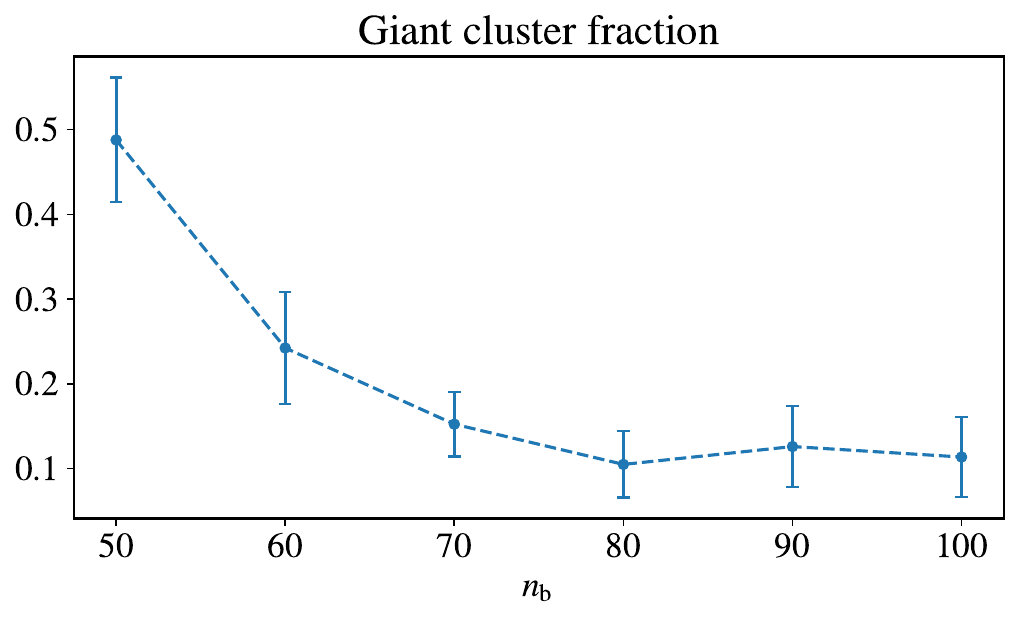}
\caption{ \textbf{Fraction of the giant cluster formed by the active Brownian particles with repulsive interactions in the presence of an upper bound on self-propulsion.} The saturation of the self-propulsion speed is given by Eq.~\eqref{eq:tanh}. The circles and the error bars indicate the average and the standard deviation, respectively, obtained using six different simulations. All parameters, except for $n_\mathrm{b}$, are identical to those used in Fig.~\ref{fig:WCA_gcf_local_density} in the main text.
}
\label{fig:tanh}
\end{figure}

In this note, we examine the case where the self-propulsion of the chemokinetic ABPs in the AMR saturates to a finite upper bound as the chemical concentration $n$ is increased. This is imposed by
\begin{align}
\label{eq:tanh}
\tilde{v}(n)=\alpha n_\mathrm{b}\tanh\left(
\frac{n}{n_\mathrm{b}}\right),
\end{align}
where $n_\mathrm{b}$ indicates the concentration scale at which $\tilde{v}(n)$ saturates, in place of $\tilde{v}(n) = \alpha n$. The same assumption was used in the literature~\cite{dinelli2023non-reciprocity} to account for the maximum self-propulsion achievable even when the chemical is abundant. In Fig.~\ref{fig:tanh}, we show how the giant cluster formed by the ABPs change as a function of $n_\mathrm{b}$. When $n_\mathrm{b}$ is reduced, the giant cluster becomes larger, as the self-propulsion increases within the clusters to a much less degree. In contrast, as $n_\mathrm{b}$ is increased, the upper limit on the self-propulsion is relaxed, making $\tilde{v}(n) \simeq \alpha n$ a better approximation. Thus, the monotonic decrease of the giant cluster fraction with increasing $n_\mathrm{b}$ confirms our view that the chemokinetic speedup inside the cluster is responsible for the supression of MIPS in the AMR.

\begin{figure}
\includegraphics[width=0.75\columnwidth]{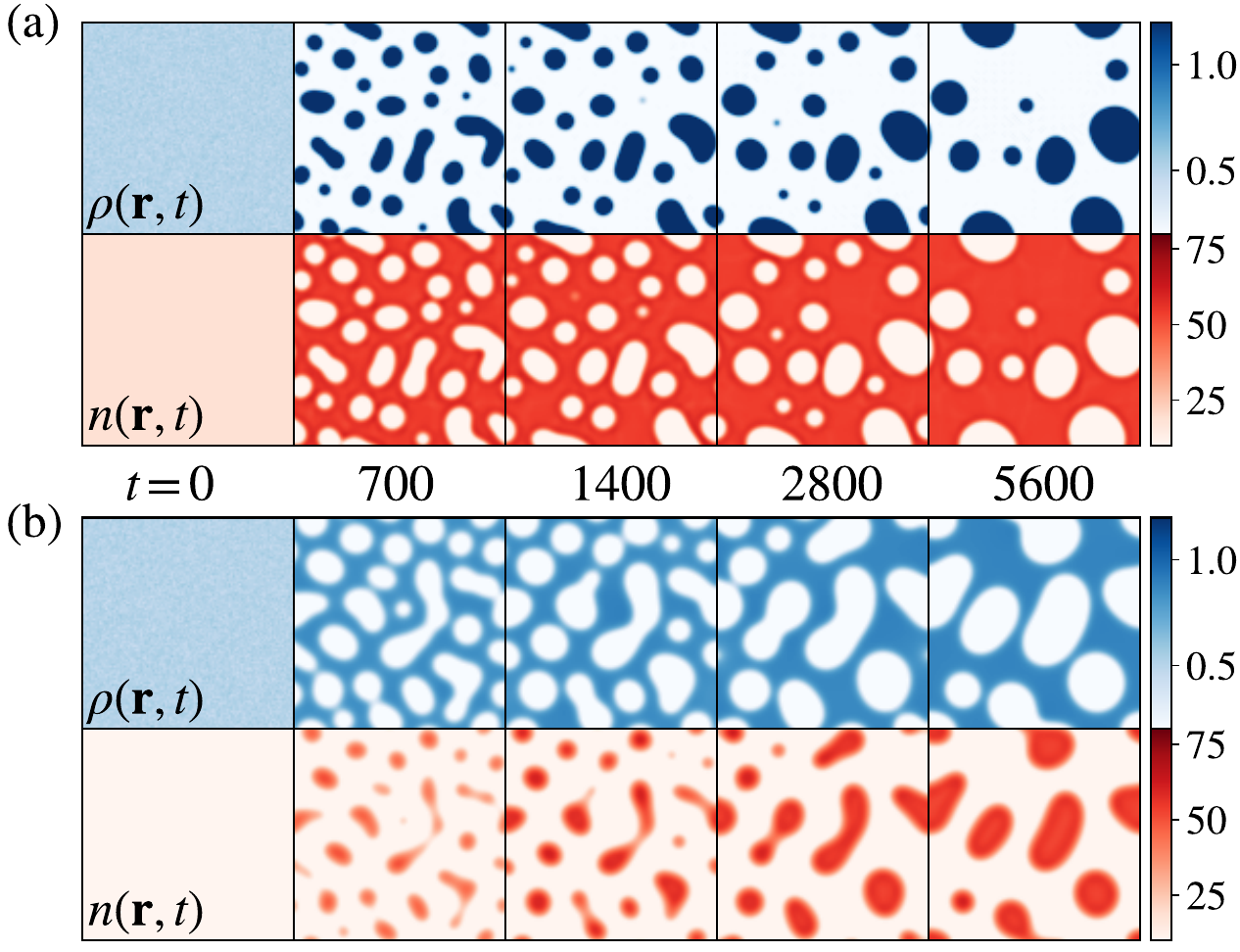}
\caption{ 
\textbf{Evolution of particle density $\rho(\mathbf{r},t)$ and chemical concentration $n(\mathbf{r},t)$ in the Basal Metabolic Regime of the hydrodynamic model.} We use (a) $n_0=18$, $\zeta = 6$ and (b) $n_0 = 6$, $\zeta = 6$, while the other parameters are identical to those used in Fig.~3(a) in the main text. Even as the global particle density remains the same ($\rho_0 = 1/2$), the system can be made to form either (a) clusters or (b) bubbles by changing the global chemical concentration $n_0$. See Supplementary Movies 5A and 5B for animated versions of (a) and (b), respectively.}
\label{fig:hydro_BMR_time_evol}
\end{figure}

\newpage
\section{Evolution of patterns in hydrodynamic simulations}
\label{sec:AppendixD}
In this note, we present demonstrations of how the particle density and the chemical concentration of the hydrodynamic model evolve to form clusters or bubbles. We first show how the phase-separated patterns in the BMR evolve with time, see Fig.~\ref{fig:hydro_BMR_time_evol}. For the vanilla MIPS, whether phase separation occurs via bubble or cluster formation depends on the global density $\rho_0$ of particles. However, by introducing chemokinetic effects, the particles can be made to form clusters [as shown in Fig.~\ref{fig:hydro_BMR_time_evol}(a)] or bubbles [as shown in Fig.~\ref{fig:hydro_BMR_time_evol}(b)] even at the same value of $\rho_0$ by adjusting the global chemical concentration $n_0$. This is thanks to the strong anticorrelation between $\rho(\mathbf{r},t)$ and $n(\mathbf{r},t)$, which makes the chemicals form clusters (bubbles) where the particles form bubbles (clusters). Since large (small) $n_0$ favors bubble (cluster) formation of the chemicals, the particles are driven to form clusters (bubbles). We also note that the length scale of the bubbles and clusters gradually increases in time, which suggests the coarsening dynamics driven by the Ostwald process.

\begin{figure}
\includegraphics[width=0.75\columnwidth]{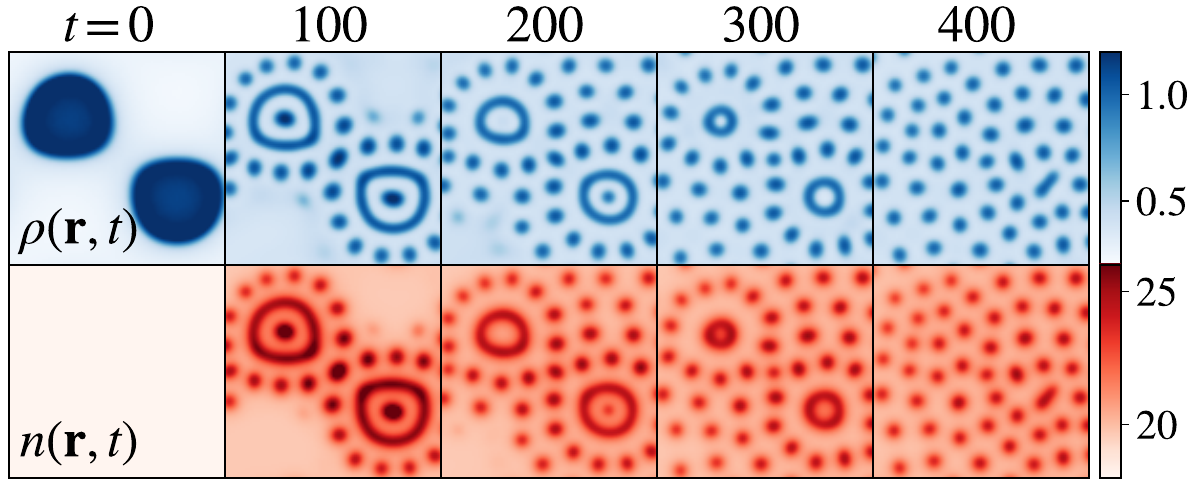}
    \caption{\label{fig:AMR_time_evol} 
    \textbf{Evolution of particle density $\rho(\mathbf{r},t)$ and chemical concentration $n(\mathbf{r},t)$ in the hydrodynamic simulation of the Active Metabolic Regime, starting from the initial state with two particle clusters and the uniform concentration.}
    The parameters are given by $n_0=20$, $\zeta=25$, $D_\mathrm{c} = 0.1$, and $\lambda = 0.5$, which corresponds to the S phase. See Supplementary Movie 6 for an animated version.}
\end{figure}

Next, we turn to the AMR and examine how the microphase separation occurs in the S phase, starting from a configuration of two large particle clusters and the uniform chemical distribution. As shown in Fig.~\ref{fig:AMR_time_evol}, these large particle clusters quickly disintegrate into smaller clusters (including some ring-like structures), while the chemical develops a pattern that overlaps almost exactly with the particle distribution. Then these intermediate particle--chemical clusters gradually shrink, letting the other particle--chemical clusters grow at their expense until all clusters have roughly the same size. This suggests the reverse Ostwald process underlying the microphase separation, which was also found to play a crucial role in the microphase separation of generic single-species scalar active matter~\cite{Tjhung2018}.

\newpage
\section{Extra results for quorum-sensing active particles}
\label{sec:AppendixE}

\begin{figure}
\includegraphics[width=0.6\columnwidth]
{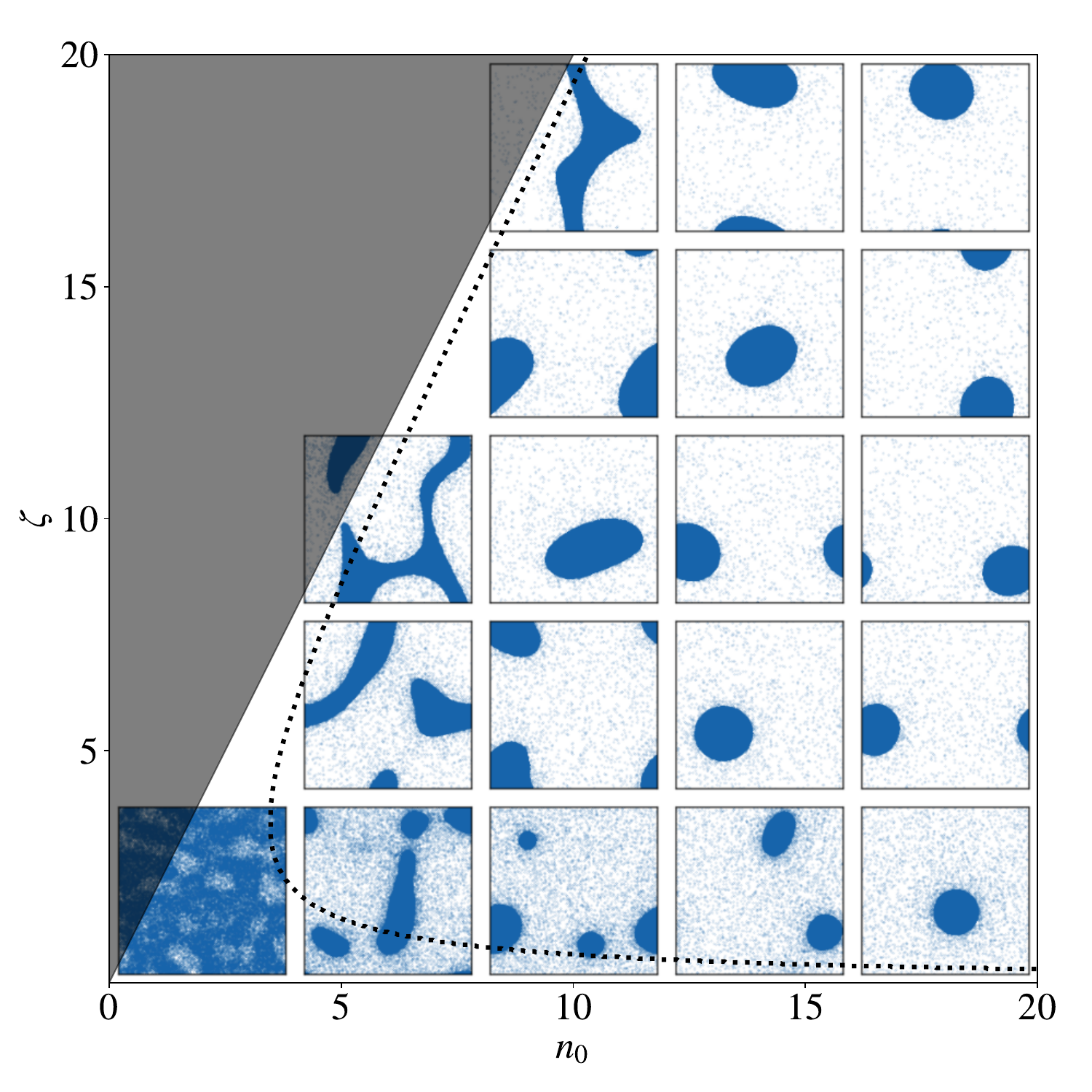}
    \caption{
    \textbf{Phase diagram of the quorum-sensing active Brownian particles in the Basal Metabolic Regime}. We use $\lambda = 0.5$ and $D_\mathrm{c}=10$, with the system size fixed at $L = 400$. The LSA prediction for the H--O boundary is indicated by the dotted curve. The snapshots of particle configurations are taken at $t = 2\times 10^3$, with the values of $\zeta$ and $n_0$ given by the coordinates of each snapshot's center.}
    \label{fig:appendix_qs_BMR}
\end{figure}

\begin{figure}
\includegraphics[width=0.75\columnwidth]
{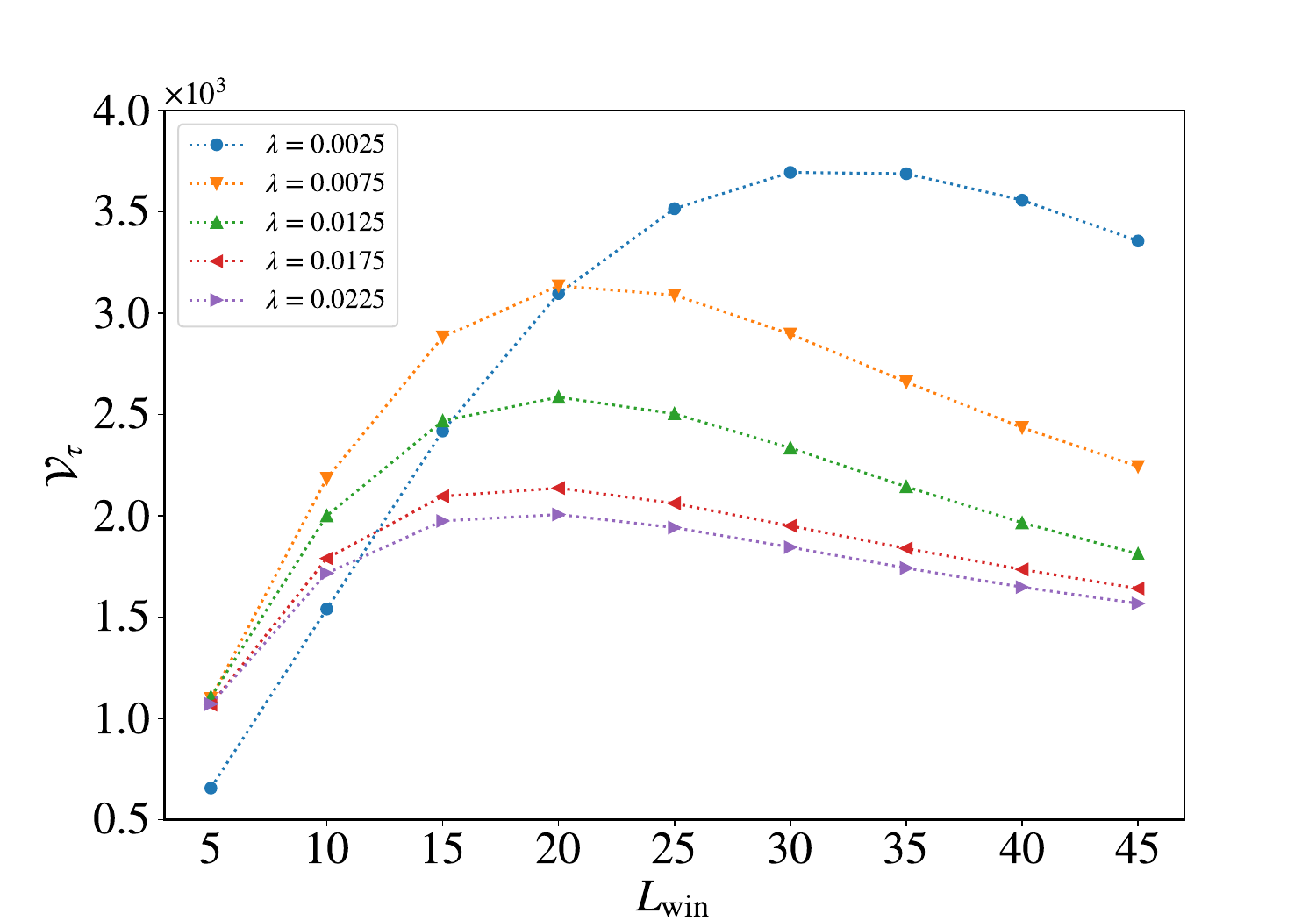}
    \caption{
    \textbf{Time-variability of the quorum-sensing active Brownian particles in the Active Metabolic Regime.} We fix $\tau = 200$ and $D_\mathrm{c} = 0.25$ as $\lambda$ and $L_\mathrm{win}$ are varied. The other parameters are identical to those used in Fig.~6(b) in the main text.}
    \label{fig:appendix_qs_AMR}
\end{figure}

In this note, we present extra results for quorum-sensing ABPs. In Fig.~\ref{fig:QS_phase_diagrams}, we show the patterns to which the clusters formed by the quorum-sensing ABPs in Fig.~\ref{eq:hydro}(a) in the main text in evolve if we run the simulations to $t = 2 \times 10^3$. We observe that the small particle clusters merge via the Ostwald process to form large-scale clusters, in agreement with the hydrodynamic prediction. Also note that the particle clusters tend to become elongated as $n_0$ becomes smaller, which is also observed in the hydrodynamic model.

In Fig.~\ref{fig:appendix_qs_AMR}, we show how the time-variability $\mathcal{V}_\tau$ depends on the window size $L_\mathrm{win}$ for various values of $\lambda$ along the $D_\mathrm{c} = 0.25$ line of Fig.~\ref{fig:QS_phase_diagrams}(b) in the main text. We note that the value of $L_\mathrm{win}$ at which $\mathcal{V}_\tau$ is maximized can be regarded as the length scale of oscillating patterns. Then, Fig.~\ref{fig:appendix_qs_AMR} indicates that oscillations occur at a shorter length scale with smaller amplitudes as $\lambda$ grows. This is consistent with the system crossing over from the O phase at small $\lambda$ to the H phase at large $\lambda$.

\end{document}